\documentclass[3p]{elsarticle}

\usepackage{lineno,hyperref}
\usepackage{commath} 
\usepackage{booktabs, longtable} 
\usepackage{siunitx} 
\usepackage{subcaption}   
\usepackage{amsmath}

\newcommand{\ars}[1]{\renewcommand{\arraystretch}{#1}}   

\modulolinenumbers[5]

\journal{Computational Materials Science}

\begin{document}

\begin{frontmatter}

\title{Nano-patterning of surfaces by ion sputtering:
Numerical study of the anisotropic damped Kuramoto-Sivashinsky equation}

\author[uerj]{E. Vitral\corref{correspondingauthor}}
\cortext[correspondingauthor]{Corresponding author}
\ead{eduardo.vitral@gmail.com}

\author[ubi]{D. Walgraef}

\author[uerj]{J. Pontes}

\author[uerj]{G. R. Anjos}

\author[uerj]{N. Mangiavacchi}

\address[uerj]{GESAR, Department of Mechanical Engineering, State University of Rio de Janeiro, Rio de Janeiro, RJ, 20940-903, Brazil}
\address[ubi]{IFISC, University of Balearic Islands, Palma de Mallorca, Islas Baleares, 07122, Spain}

\begin{abstract}
	Nonlinear models for pattern evolution by ion beam sputtering on a material surface
 	present an ongoing oportunity for new numerical simulations. A numerical analysis of
 	the evolution of preexisting patterns is proposed to investigate surface dynamics, based 
 	on a 2D anisotropic damped Kuramoto-Sivashinsky equation, with periodic boundary 
 	conditions.  A finite-difference semi-implicit time splitting scheme is employed on the 
 	discretization of the governing equation. Simulations were conducted with realistic 
 	coefficients related to physical parameters (anisotropies, beam orientation, diffusion). 
 	The stability of the numerical scheme is analyzed with time step and grid spacing tests 
 	for the pattern evolution, and the Method of Manufactured Solutions has been used to 
 	verify the proposed scheme. Ripples and hexagonal patterns were obtained from a 
 	monomodal initial condition for certain values of the damping coefficient, while 
 	spatiotemporal chaos appeared for lower values. The anisotropy effects on pattern 
 	formation were studied, varying the angle of incidence of the ion beam with respect to
 	the irradiated surface. Analytical discussions are based on linear and weakly nonlinear analysis.
\end{abstract}

\begin{keyword}
	Ion beam sputtering \sep Damped Kuramoto-Sivashinsky 
	\sep Method of Manufactured Solutions
\end{keyword}

\end{frontmatter}



\section{Introduction}

    The present endeavor is interested in the spontaneous well-ordered periodicity
    developed by ion beam sputtering (IBS), which appears over a broad surface area
    under certain conditions \cite{chason2010spontaneous}. Sputtering can be
    described as the ejection of atoms from a solid surface, as a result of energetic
    particle incidence. Among other contemporary techniques in materials science,
    sputtering operates in nonequilibrium conditions, allowing the processing of
    nano-structures beyond the limitations imposed by equilibrium thermodynamics.
    Modeling the nonlinear evolution of sputter-eroded surfaces is an ongoing 
	mechanical challenge. Our effort aims toward the development of a numerical
    scheme to solve an anisotropic Kuramoto-Sivashisnky equation with realistic
    coefficients, since it produces a rich zoology that can be adjusted to represent
    the aforementioned erosion dynamics.

	When the ion reaches the surface with a certain level of energy, a train of
    collision events may be established, resulting in the removal of atoms from that
    solid surface. The morphology of such surface can drastically change due to these
    sputtered atoms, and it might result in the formation of unexpectedly organized
    patterns, such as ripples, nanodots and hexagonal arrays of nanoholes (see Refs.
    \cite{gago2006temperature, wei2009self} for more details). Valbusa et al.
    \cite{valbusa2002nanostructuring} discussed the interplay between ion erosion and
    vacancies on the surface reorganization, which would explain some of the patterns
    experimentally detected. The rate of energy deposition is a crucial parameter for 
	the mechanisms, since high values can lead to a local transient melting of the
    surface \cite{mollick2014formation}, alongside the possibility of ion
    implantation.

	The Kuramoto-Sivashinsky equation appears from the continuum theories, in their 
	attempt to describe surfaces eroded by ion bombardment, which would ultimately 
	reproduce ripple formation and other organized patterns behavior. This equation
    was initially formulated to describe flame fronts and chemical waves 
	\cite{makeev2002morphology}, being capable of producing a great variety of 
	morphologies for its highly nonlinear and deterministic character. Later it
    became a paradigm for pattern formation and spatiotemporal chaos, with a number
    of works devoted to it. Rost and Krug \cite{rost1995anisotropic} describe the
    equation as being remarkable for the stabilization of the linear instability by
    the nonlinear term. This stabilization makes the equation a good candidate to
    represent the complexity behind the structure formation on sputtered surfaces,
    with a dynamic transiting between different regimes.

    The isotropic damped Kuramoto-Sivashinsky (DKS) was previously studied by Paniconi
    and Elder \cite{paniconi1997stationary}. They numerically integrated the 2D
    equation through Euler's method (explicit) with a time step $\Delta t = 0.035$,
    since their aim was at a late-time long-aspect-ratio limit. For the spatial
    discretization with $\Delta x = 1.0$, a midpoint rule was adopted for the first
    order derivatives, and a stencil was used for the isotropic discrete Laplacian.
    Results obtained when changing the damping coefficient were analyzed, but the
    model was not physically linked to any particular phenomenon. Facsko et al.
    \cite{facsko2004dissipative} became interested in the 2D isotropic DKS due
    to its stationary solutions showing a remarkable resemblance to IBS patterns.
    They connected these solutions to the hexagonal patterns found on a GaSb(100)
    surface after ion erosion. In terms of numerics, they also used Euler's method
    for time integration and spatially discretized the Laplace operator through
    eight nearest neighbors, and the spatial and time steps where $\Delta x = 1.0$
    and $\Delta t = 0.01$.
    
    Here we perform a numerical study of patterning by IBS through a model
    that corresponds to a DKS equation for an anisotropic system with realistic
    coefficients. Our proposed model is close to a previous expansion of the Bradley
    and Harper theory performed by Makeev et al. \cite{makeev2002morphology}, and
    also to the form found in Facsko and Keller \cite{keller2010ion}, but contains
    terms to account for various anisotropies and adopts some simplifications (as
    isotropic energy distribution for the atomic cascade). Since the equation is very
    sensitive to its parameters value, there is still much to be explored when it
    comes to its behavior in 2D, especially when working in the range of physical
    experimental data. Our goal is to understand the evolution of a material surface
    displaying a preexisting pattern under IBS through this model. We look into the
    growth and competition between modes for different initial patterns, which are
    compared with analytical results from a developed linear and weakly nonlinear
    analysis. Also, we study the connection between the beam angle and the
    relative anisotropy found on the patterns, and provide an insight on the
    role of the damping term under our particular model and coefficients.
    This way, we introduce tools and a new way to interpret experimental
    results, from ripples to coarsening in the nonlinear regime.
    
    Stability requirements impose restrictive limitations to the time
    step on explicit schemes, especially in the presence of fourth-order derivatives.
    At the same time, implementation of straight implicit schemes for 2 or more
    dimensions leads to a large system of linear equations, which might not be a
    suitable cost-efficient option. Therefore, in Sec. \ref{sec:model} we propose a
    finite-difference semi-implicit splitting scheme of second order in time and
    space to numerically solve this anisotropic DKS equation subjected to periodic
    boundary conditions. The computational domain is a two dimensional surface
    characterized by a height function $h(x,y,t)$, whose evolution in time is
    monitored. Internal iterations are used inside each time step to enhance the
    approximation of the nonlinear term.

    Previously, a similar numerical scheme has been successfully implemented for
    Swift-Hohenberg \cite{christov1997implicit, christov2002numerical}, which is a
    fourth-order parabolic equation, dealing with the same challenges of high-order
    spatial derivatives and nonlinearity. The coordinate splitting, alongside the
    half-time steps (with only one of the operators implemented explicitly), were
    proven to be an effective approach for such equation, combining desirable
    stability properties with efficient computational costs. We show that this is
    also true for our case, such that the scheme remains stable for time steps much
    larger than the reported ones for explicit methods.

    Section \ref{sec:linear} deals with a linear analysis to study the system's
    response to small perturbations, followed by a weakly nonlinear analysis in
    Sec. \ref{sec:wna}. The numerical scheme is verified by the Method of
    Manufactured Solutions (MMS) in Sec. \ref{sec:verification}, and its stability
    is analyzed in Sec. \ref{sec:stability} with time step and grid spacing tests
    for the pattern evolution. Preexisting structures are employed as initial
    conditions, varying from a monomodal to a random initial pattern.

    We then further discuss how the realistic coefficients fit in the model in Section
    \ref{sec:dks}, where simulation results are shown for a high temperature case.
    In such system, sputtering terms and diffusion act on similar scale, and
    should agree with our qualitative analysis for pattern selection. There,
    we study the circumstances under which hexagonal patterns emerge, and how the
    damping coefficient $\alpha$ relates to the organizational degree of the
    structure (even for spatiotemporal chaos). Finally, we analyze the effects of
    variations of the angle of incidence of the beam $\theta$, and also a scenario
    where the nonlinear terms compensate each other.


\section{Mathematical Modelling and Numerical Scheme}
\label{sec:model}

	\subsection{Governing Equation}

	In order to solve the anisotropic DKS equation which emerges from the IBS
    modelling \cite{valbusa2002nanostructuring,makeev2002morphology,sigmund1969theory,
    bradley1988theory,cuerno1995dynamic}, a second order in time finite difference
    numerical scheme is proposed. The general model takes into account 
	realistic coefficients corresponding to anisotropies, diffusion, beam orientation
    and others. For the case of isotropic energy distribution, considering an ion beam
    with angle of incidence $\theta$ with respect to the normal of the surface
    ($\theta = 0$ for normal incidence), the evolution of the surface height $h$
    is governed by:

    \begin{eqnarray}
      \nonumber
      \frac{\partial h}{\partial t}
      &=&
      -\alpha_o\, h + \frac{F a}{2}\left(\mu \frac{\partial^2 h}{\partial x^2}
      + \nu \frac{\partial^2 h}{\partial x^2}\right)
      +\frac{F a^2_\eta}{2}\left(\nu_x\left(\frac{\partial h}{\partial x}\right)^2
      +\nu_y \left(\frac{\partial h}{\partial y}\right)^2  \right)
      \\[3mm]
      \nonumber
      &&
      +\frac{Fa^3}{8a^2_\eta}\left(D_{XX}\frac{\partial^4 h}{\partial x^4}
      +D_{XY}\frac{\partial^4 h}{\partial x^2y^2}
      +D_{YY}\frac{\partial^4 h}{\partial y^4}\right)
      \\[3mm]
      \label{eq:KS1}
      &&
      -K\left(\frac{\partial^4 h}{\partial x^4}
      +2\frac{\partial^4 h}{\partial x^2y^2}
      +\frac{\partial^4 h}{\partial y^4}\right) \,.
    \end{eqnarray}
    
    Here, $F = \frac{J\epsilon p}{\sqrt{2\pi}\eta}{\rm exp}
    \left(\frac{-a^2_\eta c^2}{2}\right)$, where $J$ is the flux of bombarding ions,
    $\epsilon$ is the energy carried by the ions, p is associated to the surface
    binding energy and scattering cross-section, $a$ is the penetration depth, and
    $\eta$ is the width of energy distribution. The parameter $K$ takes into account
    the surface diffusion effects, which vary with temperature. In terms of realistic
    values, $\epsilon$ lies between 0.1 and 100 keV, while some examples for the
    others are:
    $J \approx 10^{15} {\rm cm}^{-2}{\rm s}^{-1}$, $p \approx 2$, $a \approx 2$ nm,
    $\eta \approx 0.5$ nm, and $K \approx 34 \times 10^{-28}{\rm cm}^4{\rm s}^{-1}$
    \cite{makeev2002morphology,keller2010ion}.
    Additionally, $c$ and $s$ represent the cosine and
    sine of $\theta$, and $a_\eta$ is the ratio $a/\eta$. In standard form,
    the equation reads:

	\begin{eqnarray}
		\nonumber
		\frac{\partial\bar h}{\partial \tau}
		&=&
		-\alpha\,\bar h+\mu\frac{\partial^2\bar h}{\partial X^2}
		+ \nu \frac{\partial^2\bar h}{\partial Y^2}
		+\bar\nu_x\bigg(\frac{\partial\bar h}{\partial X}\bigg)^2
		+\bar\nu_y\bigg(\frac{\partial\bar h}{\partial Y}\bigg)^2
		+D_{XX}\frac{\partial^4\bar h}{\partial X^4}
		\\[3mm]
		\label{eq:KS2}
		&&
		+D_{XY}\frac{\partial^4\bar h}{\partial X^2\partial Y^2}
		+D_{YY}\frac{\partial^4\bar h}{\partial Y^4}
		-\bar K\bigg(
			\frac{\partial^4\bar h}{\partial X^4}
			+2\frac{\partial^4\bar h}{\partial X^2Y^2}
			+\frac{\partial^4\bar h}{\partial Y^4}
		\bigg)
	\end{eqnarray}

	\noindent where $\bar{h}$ and $\tau$ are, respectively, the dimensionless
    surface height function of the external atom layer and the time dependency of
    the transient model, with $X$ and $Y$ as the domain space coordinates. They
    relate to their dimensional counterpart via: $\bar{h} = \frac{a^2_\eta}{a}h$,
    $X = \frac{2 a_\eta}{a} x$ (similarly for Y), $\tau = \frac{2 F a^2_\eta}{a} t$,
    $\alpha = \frac{a}{2Fa^2_\eta}\alpha_o$, and $\bar{K} = \frac{8a^2_\eta}{Fa^3}K$.
    Equation \ref{eq:KS2} presents a damping term $-\alpha\,\bar{h}$ , with 
	$\alpha$ being a positive damping coefficient, which was initially proposed as
    a contribution of the redeposition mechanism to the formation of nanodots
    \cite{facsko2004dissipative}. However, Bradley \cite{bradley2011redeposition}
    demonstrated that the redeposition of sputtered material is a nonlinear effect
    observed in pattern formation by IBS, discarding the influence of this physical
    mechanism on the appearance of hexagonal ordered structures. Nevertheless, we
    maintain our interest in the linear damping term for the sake of producing
    stable patterns and supressing the emergence of chaos . Finally, the
    parameters $\mu$,  $\bar \nu_x$, $\bar \nu_y$, $D_{XX}$ and
    $D_{XY}$ will be defined as follows (see Makeev et al.
    \cite{makeev2002morphology} for further details):
	
	\vspace{-5mm}
	
	\begin{flalign*}
		& \hspace{5mm}\mu   = 2s^2-c^2-a^2_\eta s^2c^2 &\nu = -c^2  \hspace{15mm}
		\\[3mm]
		&\hspace{5mm}\bar\nu_x = c\,\bigg(3s^2-c^2-a_\eta^2s^2c^2\bigg)  &\bar\nu_y = -c^3  \hspace{15mm}
		\\[3mm]
		&\hspace{5mm}D_{XX} =
		- \bigg(
			c^2-4s^2+2a^2_\eta s^2
			\left(c^2-\frac{2}{3}s^2\right)+
			\frac{a^4_\eta}{3}s^4c^2
		\bigg) & D_{YY} = c^2   \hspace{15mm}
		\\[3mm]
		&\hspace{5mm}D_{XY} = 2\,\bigg(c^2-2s^2+a^2_\eta s^2c^2\bigg) \;\;. &
	\end{flalign*}

	These parameters are responsible for 
	introducing various types of anisotropies. Now, in order to solve Eq. \ref{eq:KS2}, the 
	following second order in time Crank-Nicolson semi-implicit scheme was adopted:

	\begin{eqnarray}
		\label{eq:target scheme}
		\frac{\bar h^{\;n+1}-\bar h^{\;n}}{\Delta\tau}
		&=&
		\Lambda_X\left(\frac{\bar h^{\;n+1}+\bar h^{\;n}}{2}\right)
		+
		\Lambda_Y\left(\frac{\bar h^{\;n+1}+\bar h^{\;n}}{2}\right)
		+
		f^{\;n+1/2} \;.
	\end{eqnarray}

	The superscript $(n+1)$ refers to the current time and $(n)$ to the previous one. The
	operators $\Lambda_X$, $\Lambda_Y$ (both modified from Eq. \ref{eq:target scheme} to 
	account for the division by two) and the function $f^{\;n+1/2}$ are defined as:

	\begin{eqnarray*}
		\Lambda_X &=&
		\frac{1}{2}
		\left[
			-\frac{\alpha}{2}
			-\left(D_{XX}+K\right)\frac{\partial^4}{\partial X^4}
		\right]
		\\[3mm]
		\hspace*{-10mm}
		\Lambda_Y &=&
		\frac{1}{2}
		\left[
			-\frac{\alpha}{2}
			-K\frac{\partial^4}{\partial Y^4}
		\right]
		\\[3mm]
		f^{n+1/2} &=&
		\frac{1}{2}
		\left[
			\bar\nu_x
			\left(
				\frac{\partial\bar h^{\;n+1}}{\partial X}
				+\frac{\partial\bar h^{\;n}}{\partial X}
			\right)
			\frac{\partial}{\partial X}
			+\bar\nu_y
			\left(
				\frac{\partial\bar h^{\;n+1}}{\partial Y}
				+\frac{\partial\bar h^{\;n}}{\partial Y}
			\right)
			\frac{\partial}{\partial Y}
		+\mu\frac{\partial^2}{\partial X^2}
		\right.
		\\[3mm]
		&&
		\left.
		+\nu\frac{\partial^2}{\partial Y^2}
			+\left(D_{XY}-2K\right)
			\frac{\partial^4}{\partial X^2\partial Y^2}
			+D_{YY}\frac{\partial^4}{\partial Y^4}
		\right]\left(\bar h^{\;n+1}+\bar h^{\;n}\right) \;.
	\end{eqnarray*}

	\subsection{Internal Iterations}

	Internal iterations at each time step are required to secure the approximation for the
	nonlinearities taking part in the scheme of Eq. \ref{eq:target scheme}. The iterations 
	loop will continue until the $L_{\infty}$ norm points that the convergence was attained. 
	There is a trade-off related to the time step $\Delta \tau$: for a larger $\Delta \tau$, 
	convergence will be impaired and the number of internal iterations will increase, while 
	a smaller $\Delta \tau$ will impact on a smaller number of iterations, but it will imply 
	on a greater number of time steps. The internal iterations scheme reads:

	\begin{eqnarray}
		\label{eq:internal}
		\frac{\bar h^{\;n,m+1}-\bar h^{\;n}}{\Delta\tau}
		&=&
		\Lambda_X\left(\bar h^{\;n,m+1}+\bar h^{\;n}\right)
		+ \Lambda_Y\left(\bar h^{\;n,m+1}+\bar h^{\;n}\right)
		+ f^{\;n+1/2}
	\end{eqnarray}

	\noindent where the index $(m)$ refers to the internal iteration number. The superscript 
	$(n,m + 1)$ identifies the new iteration, while $(n)$ are the values of the previous time 
	step. The superscript $(n+1)$ for the nonlinear term in the function $f^{n+1/2}$ will be 
	replaced by $(n,m)$, which stands for the values obtained from the previous iteration. 
	The iterations proceed until the following criterion for the $L_{\infty}$ norm is 
	satisfied:


	\begin{eqnarray}
		L_\infty &=& 
		\frac{\textit{max}\;\abs{\bar h^{\;n,m+1}-\bar h^{\;n,m}}}
		{\raisebox{-0.75ex}{$max\;\abs{\bar h^{\;n,m+1}}$}} \; < \; 10^{-7}
	\end{eqnarray}

	\noindent for a fixed current time $(n)$. The function $\bar{h}^{\;n+1}$ for the 
	new time will be acquired from $\bar{h}^{\;n,m+1}$, as soon as the criterion of convergence
	is satisfied.

	\subsection{The Splitting Scheme}

	The splitting of Eq. \ref{eq:target scheme} is made according to the \textit{second 
	Douglas scheme} \cite{douglas1956numerical,yanenko1971method}. Such strategy has been 
	chosen to deal with the costly procedure of solving Eq. \ref{eq:target scheme}; even 
	though we are working with sparse matrices for the operators, the internal iterations cause 
	the process to be repeated several times during each time step. This problem welcomes an 
	attempt to minimize the operations per unit iteration, as follows:

	\begin{eqnarray}
		\label{eq:splt1}
		\frac{\tilde{\bar h}-\bar h^{\;n}}{\Delta\tau}
		&=&
		\Lambda_X\,\tilde{\bar h}+\Lambda_Y\,\bar h^{\;n}+
		f^{\;n+1/2}+
		\left(\Lambda_X+\Lambda_Y\right)\bar h^{\;n}
		\\[3mm]
		\label{eq:splt2}
		\frac{\bar h^{\;n,m+1}-\tilde{\bar h}}{\Delta\tau}
		&=&
		\Lambda_Y\left(\bar h^{\;n,m+1}-\bar h^{\;n}\right) \; .
	\end{eqnarray}

	Here, $\tilde{\bar h}$ is the height function for the intermediary time step. Equation 
	\ref{eq:splt1} can be solved line by line and Eq. \ref{eq:splt2} can be solved column by 
	column, which is one positive contribution of the splitting for the storage and precision of the 
	resolution. In order to show that the splitting represents the original scheme, we rewrite 
	Eqs. \ref{eq:splt1} and \ref{eq:splt2} in the form:
	
	\begin{eqnarray}
		\label{eq:splt3}
		\hspace*{-15mm}
		\left(E-\Delta\tau\,\Lambda_X\right)\tilde{\bar h}
		&=&
		\left(E+\Delta\tau\,\Lambda_X\right)\bar h^{\;n}
		+2\Delta\tau\,\Lambda_Y\bar h^{\;n}
		+\Delta\tau\,f^{\;n+1/2}
		\\[3mm]
		\label{eq:splt4}
		\left(E-\Delta\tau\,\Lambda_Y\right)\bar h^{\;n,m+1}
		&=&
		\tilde{\bar h}-\Delta\tau\,\Lambda_Y\bar h^{\;n}
	\end{eqnarray}
	
	\noindent where $E$ is the identity operator. The intermediate variable $\tilde{\bar h}$ is 
	eliminated by applying the operator $\left(E-\Delta \tau\,\Lambda_X\right)$ to Eq. 
	\ref{eq:splt4}, and adding the result to Eq. \ref{eq:splt3}:
	
	\begin{eqnarray*}
		\left(E-\Delta\tau\,\Lambda_X\right)
		\left(E-\Delta\tau\,\Lambda_Y\right)\bar h^{\;n,m+1}
		&=&
		\left(E+\Delta\tau\,\Lambda_X\right)\bar h^{\;n}+
		2\Delta\tau\,\Lambda_Y\bar h^{\;n}
		\\[3mm]
		&&
		+\Delta\tau\,f^{\;n+1/2}		
		-\left(E-\Delta\tau\,\Lambda_X\right)\Delta\tau\,\Lambda_Y\bar h^{\;n} \; .
	\end{eqnarray*}
	
	This result may be rewritten as:
	
	\begin{eqnarray*}
		\left[
			E-\Delta\tau\left(\Lambda_X+\Lambda_Y\right)
			+\Delta \tau^2\,\Lambda_X\Lambda_Y
		\right]
		\bar h^{\;n,m+1}
		\;\; =
		\\[3mm] && \hspace{-40mm}
		\left[
			E+\Delta\tau
			\left(\Lambda_X+\Lambda_Y\right)
			+\Delta \tau^2\,\Lambda_X\Lambda_Y
		\right]
		\bar h^n
		+\Delta \tau\,f^{\;n+1/2}
	\end{eqnarray*}
	
	or either:
	
	\begin{eqnarray}
		\nonumber
		\left(E+\Delta \tau^2\,\Lambda_X\Lambda_Y\right)
		\frac{\bar h^{\;n,m+1}-\bar h^{\;n}}{\Delta \tau}
		\;\; =
		\\[3mm] && \hspace{-15mm} \label{eq:equiv}
		\left(\Lambda_X+\Lambda_Y\right)
		\left(\bar h^{\;n,m+1}+\bar h^{\;n}\right)+f^{\;n+1/2} \; .
	\end{eqnarray}
	
	A comparison with Eq. \ref{eq:target scheme} shows that Eq. \ref{eq:equiv} is 
	actually equivalent to the original one, as their temporal order of approximation coincides.  
	Note that the positive definite operator
	
	\begin{displaymath}
		B \;\; \equiv \;\; E+\Delta\tau^2\,\Lambda_X\Lambda_Y \;\;=\;\;
		E+{\cal O}\left(\Delta \tau^2\right)
	\end{displaymath}
	
	\noindent has a norm greater than one, acting on the discrete time derivative. This means that
	 the operator  $B$ does not change the steady state solution. Furthermore, since 
	 $\norm{B}>1$, the scheme given by Eqs. \ref{eq:splt1}-\ref{eq:splt2} is more stable than the 
	 target one (Eq. \ref{eq:target scheme}).

	\subsection{The Method of Manufactured Solutions}

	The MMS is a code verification procedure, which analyzes if a numerical scheme
	and its implementation code stand for the task of representing the mathematical
	model of a physical event with sufficient accuracy. The idea behind the MMS is to
	solve a problem as if the analytical solution was available from start, creating a
	manufactured solution for a system of partial differential equations \cite{malaya2013masa}.
	Considering that the proposed function is unlikely to solve the equations exactly, a residual 
	term will appear due to the solution of the system.  The insertion of such residue in the
	right-hand side of the equation as a source term leads to a different numerical solution,
	which is expected to approach the artificial analytical solution (if the
	manufactured equation was properly constructed). 
	
	Since the manufactured solution is defined on the continuum, the global discretization error 
	can be examined by the discrete $L_2$ norm \cite{roy2005review}:

	\begin{eqnarray}
	L_{2} &=& 
	\left(\rule{0cm}{0.8cm}\right.\frac{\sum\limits_{i=1}^N \;\abs{ \bar{h}^{\;n}_{\;i,k} - \bar{h}^{\;n}_{\;i,e}}^{2}}
	{\raisebox{-0.75ex}{$N$}} \left.\rule{0cm}{0.8cm}\right)^{1/2}
	\end{eqnarray}

	\noindent where $N$ is the total number of mesh nodes, $i$ is the index of each node, 
	and the indexes $k$ and $e$ represent the numerical and the manufactured (``exact'') solution, 
	respectively. Such norm analyzes how the numerical solution approaches its corresponding
	analytical solution after each time step. It is indeed expected that the error will
	decrease by refining the mesh.


\section{Linear Stability Analysis}
\label{sec:linear}

	 Linear stability analysis is employed to study the system's behavior when submitted to
	 small perturbations, revealing whether the initial equilibrium point is stable or not. If an 
	 exponential growth is observed, such point is linearly unstable; on the other hand, if a 
	 exponential decay towards a steady state is found, that point is  classified as linearly stable. 
	 For small perturbations, we may eliminate the non-linear terms from the governing 
	 equation, making the analytical analysis more straightforward. Equation \ref{eq:KS2} will 
	 take the form:
	 
	 \begin{eqnarray}
		\nonumber
		\frac{\partial\bar h}{\partial \tau}
		&=&
		-\alpha\bar h+\mu\frac{\partial^2\bar h}{\partial X^2}
		+\bar{\nu}\frac{\partial^2\bar h}{\partial Y^2}
		-D_{XX}\frac{\partial^4\bar h}{\partial X^4}
		+D_{XY}\frac{\partial^4\bar h}{\partial X^2\partial Y^2}
		\\[3mm]
		\label{eq:linKS2}
		&&
		+D_{YY}\frac{\partial^4\bar h}{\partial Y^4}
		-\bar K\bigg(
			\frac{\partial^4\bar h}{\partial X^4}+
			2\frac{\partial^4\bar h}{\partial X^2Y^2}+
			\frac{\partial^4\bar h}{\partial Y^4}
		\bigg) \; .
	\end{eqnarray}
	 
	 Define a standard basis $\{\vec{1}_x, \vec{1}_y\}$ for the present Cartesian coordinate 
	 system. Considering $\bar{K} > 1$, a value of $\theta$ for $|\mu| > |\nu|$ (where 
	 $\nu = -c^2$), and a perturbation from the equilibrium state $\bar{h}_o$, we may 
	 write Eq. \ref{eq:linKS2} in Fourier series:
	 
	\begin{eqnarray*}
		\bar{h}\; (X,Y,t) \;\; = \;\; \sum_k \; \bar{h}_k \;e^{\;i ( q_x X + q_y Y)} e^{\;\sigma_\tau\;t}
		\hspace{30mm}
		\\[3mm]
		\sigma_\tau \; = \;
		\big[ -\alpha+(-\mu q_x^2)+(-\nu q_y^2)
		-D_{XX}q_x^4+D_{XY}q_x^2q_y^2 + D_{YY}q_y^4 
		- \bar{K} (q_x^2 + q_y^2)^2\big]
	\end{eqnarray*}
	
	\noindent where $\vec{q}$ are the spatial modes ($|q|= \sqrt{q_x^2+q_y^2}$), and 
	$\sigma_\tau$ is the growth rate (eigenvalue). The decomposition in Fourier modes is 
	reasonable because the basis of trigonometric functions is appropriate to periodic structures. 
	Since $\nu$ is negative for any $\theta$, and $\mu$ is negative for 
	$0 < \theta < 70.1^\circ$, we continue with the absolute values $|\mu|$ and $|\nu|$.
	The anisotropic coefficients $D_{XX}$, $D_{XY}$ and $D_{YY}$ can be hidden for an easier
	manipulation of the equation, observing that $\bar{K}$ multiplies the same derivatives, as 
	follows:
	
	\begin{eqnarray}
		\label{eq:est1}
		\sigma_\tau &=&  -\alpha+|\mu|q_x^2+|\nu|q_y^2
		 - \bar{K} (q_x^2+ q_y^2)^2 \; .
	\end{eqnarray}	
		
	This is an equation for the rate of growth of the mode $\vec{q}$. Consider 
	$q^2 = q_x^2+ q_y^2$ for Eq. \ref{eq:est1}. Also, if we define the critical wavenumber 
	as $q_c^2 = |\mu|/2\bar{K}$, the following expression is obtained:
	
	\begin{eqnarray*}
		\sigma_\tau &=& ( -\alpha+|\mu|q^2-\bar{K}q^4)
		-(|\mu|-|\nu|) \; q_y^2
		\\[3mm] &=&
		\epsilon-\bar{K}(q^2-q_c^2)^2-(|\mu|-|\nu|) \; q_y^2
	\end{eqnarray*}
	
	\noindent where $\epsilon = \mu^ 2/4\bar{K} - \alpha$. On decreasing 
	$\alpha$ bellow $\mu^ 2/4\bar{K}$, a positive $\epsilon$ value is 
	obtained. From this moment on, spatial modes with $\vec{q} = \pm\, q_c \vec{1}_x$ become 
	unstable, while distancing from such values we remain in the stable domain. That is:
	
	\begin{eqnarray*}
		\sigma_\tau &=& \epsilon-\bar{K}(q_x^2-q_c^2)^2 
		\;\;\; \implies \;\;\; q_x \;\; \approx  \;\; q_c \;\;\;  (\textrm{for a small positive } \epsilon) \; .
	\end{eqnarray*}	
	
	Reducing $\alpha$ even further, the unstable domain for $\vec{1}_x$ modes expands.
	For $\alpha = \nu^ 2/4\bar{K} $ and changing modes orientation up to
	$\vec{1}_y$, we can find that modes with 
	$\vec{q} = \pm \sqrt{ \nu/2\bar{K}\phantom{\frac{1}{2}}} \vec{1}_y$ 
	become first unstable in this direction. The calculation is performed as follows:
	
	\begin{eqnarray*}
		\sigma_\tau &=&
		\dfrac{\mu^2- \nu^2}{4\bar{K}}-\bar{K}(q^2-q_c^2)^2-(|\mu|
		-|\nu|) \; q_y^2
		\\[3mm] &=&
		\dfrac{\mu^2}{4\bar{K}}-\dfrac{\nu^2}{4\bar{K}}
		-\bar{k}q_y^4+q_y^2|\nu|-\dfrac{\mu^2}{4\bar{K}}
		+q_y^2|\mu|-q_y^2|\nu|
		\\[3mm] &=&
		-\dfrac{\nu^2}{4\bar{K}}-q_y^2(\bar{K}q_y^2-|\nu|)
		\;\;\; \implies \;\;\; q_y \;\; =  \;\; \pm\, \sqrt{\frac{\nu}{2\bar{K}}} \; .
	\end{eqnarray*}
	
	From the previous analysis, we can conclude that for  $\alpha < 
	\nu^ 2/4\bar{K}$, modes for all orientations will become unstable. However,
	it must be noticed that the maximum growth rate is limited to $q_c \vec{1}_x$ , since the
	term $\bar{K}(q^2-q_c^2)^2$ acts as a wave filter.
	
	The linear stability analysis does not take into account weakly nonlinear effects. 
	A more robust analysis of the prevailing structure should consider the interactions
	between different modes and the nonlinearities of the system. This coupling can be
	studied through amplitude equations. For the hexagonal structures, the modes are of equal 
	amplitude, although presenting different wavevectors in order to deal with the anisotropy 
	\cite{ghoniem2008instabilities}.

\section{Weakly nonlinear Analysis}
\label{sec:wna}

	Based on the linear stability analysis,  for $\alpha$ in range $ \nu^ 2/4\bar{K} \leq 
	\alpha \leq \mu^ 2/4\bar{K}$ the maximum growth rate is associated with a 
	$\vec{q} = q_c \vec{1}_x$  monomode pattern. Nevertheless, when we take into
	account nonlinear effects, this pattern may become unstable versus structures built on 
	modes directly coupled through quadratic nonlinearities 
	\cite{walgraefProposal,manneville1990dissipative}. The hexagonal modes are such that:
	
	\begin{eqnarray*}
	q_c \vec{1}_x + \vec{q}_2 + \vec{q}_3 =  0
	\end{eqnarray*}

	\noindent where $q_c$ defines the critical circle $q_c = | \vec{q}_i | $, and $\vec{q}_2$ 
	and $\vec{q}_3$ are defined as: 

	\begin{eqnarray*}
	\vec{q}_2 &=& q_{2x}\,\vec{1}_x + q_{2y}\,\vec{1}_y
	\\[3mm] 
	\vec{q}_3 &=& q_{3x}\,\vec{1}_x + q_{3y}\,\vec{1}_y \; .
	\end{eqnarray*}

	Therefore, $q_{2x} = q_{3x} = - q_c / 2$ and $q_{2y} = -q_{3y} = \sqrt{3} q_c / 2 $.
	For critical hexagons, we propose the solution:
	
	\begin{eqnarray*}
	\bar h = \sum\limits_{n}( A_n e^{i\,n\,\vec{q}_1 \cdot \vec{r}}
	+ B_n e^{i\,n\,\vec{q}_2 \cdot \vec{r}} + C_n e^{i\,n\,\vec{q}_3 \cdot \vec{r}} + c.c. ) \; .
	\end{eqnarray*}
	
	Before moving to the amplitude equations, it is more straightforward to start analyzing 
	from a critical ripple. The unidimensional evolution equation is given by:
	
	\begin{eqnarray*}
	\frac{\partial \bar h}{\partial \tau}
	& = &
	- \alpha h + \mu \frac{\partial^2 h}{\partial X^2} + \bar\nu_x 
	\bigg(\frac{\partial h}{\partial X}\bigg)^2
	- \bar K \frac{\partial^4 h}{\partial X^4}
	\\[3mm]
	& = &
	\epsilon h - \bar K \bigg( q_c^2 + \frac{\partial^2}{\partial X^2} \bigg)^2 h  
	+ \bar\nu_x \bigg(\frac{\partial h}{\partial X}\bigg)^2
	\end{eqnarray*}

	\noindent where $\epsilon = \mu^2 / 4 \bar K - \alpha$, and $q_c^2 = |\mu| / 2 \bar K $ (the 
	coefficient $\mu$ is negative for the angles of incidence covered in this work). Then, if we 
	consider a solution of the form  $h = A_1 e^{i \,q_c x} + A_2 e^{i \,2  q_c  x} + ... + c.c.$, the 
	following results are obtained:
	
	\begin{flalign*}
	a)\;\;\; \bigg(\frac{\partial \bar h}{\partial X}\bigg)^2
	& = 
	(i\,q_c\,A_1 e^{i \,q_c x} + i \, 2 \, q_c \,A_2 e^{i \,2 q_c x} + ... + c.c. )^2 &
	\\[3mm]
	& = 
	( -q_c^2\,A_1^2\,e^{i \,2 q_c x}  + 4\,q_c^2\,A_2\,A_1^*\,e^{i \,q_c x}+ ... + c.c. )
	\end{flalign*}
	
	\begin{flalign*}
	&b) \;\;\;\bigg( q_c^2 + \frac{\partial^2}{\partial X^2}\bigg)^2 A_ 2\,e^{i \,2 q_c x}
	 = 
	(q_c^2 - 4\, q_c^2)^2 A_ 2\,e^{i \,2 q_c x} = -9\,q_c^4\,A_ 2\,e^{i \,2  q_c  x} \; . &
	\end{flalign*}
	
	Hence, the amplitude equations for $A_1$ and $A_2$ become:
	
	\begin{eqnarray}
	\label{eq:ampA1}
	\frac{\partial A_1}{\partial \tau} &=& \epsilon A_1 + 4\, q_c^2 \bar\nu_x A_2 A_1^* + ...
	\\[3mm]
	\label{eq:ampA2}
	\frac{\partial A_2}{\partial \tau} &=& (\epsilon - 9\, \bar K q_c^4) A_2  - q_c^2 \bar\nu_x A_1^2 + ...
	\end{eqnarray}
	
	For slow variations of the amplitude, the time derivative in Eq. \ref{eq:ampA2} becomes
	small when compared to the first two terms in the right-hand side such that we may write:
	
	\begin{eqnarray*}
	A_2 &\approx& \frac{q_c^2 \bar\nu_x A_1^2}{\epsilon - 9\, \bar K q_c^4} \; .
	\end{eqnarray*}

	Hence, the amplitude $A_2$ adiabatically follows the amplitude $A_1$ as:

	\begin{eqnarray*}
	\frac{\partial A_1}{\partial \tau} &=& \epsilon A_1 - u |A_1|^2 A_1 + ... \; , \hspace{5mm}
	\textrm{with} \;\; u  \approx \frac{4\, \bar\nu_x^2}{\,9 \bar K} \;\; \textrm{for small} \; \epsilon .
	\end{eqnarray*}

	Now, we move back to the critical hexagons. The evolution of $A_1$ behaves similarly
	to the critical ripples, and the only difference in the equation comes from the
	interaction between the second and the third modes. Note that the following terms are
	obtained:

	\begin{flalign*}
	a)\;\;\; \bigg(\frac{\partial\bar h}{\partial X}\bigg)^2
	& =  
	-q_c^2\,A_1^2\,e^{i \,2 q_c x}  + 4\,q_c^2\,A_2\,A_1^*\,e^{i \,q_c x}
	- \frac{1}{2}\, q_c^2 B_1^* C_1^* \epsilon^{i \,q_c x}
	 +... + c.c. &
	\end{flalign*}
	
	\begin{flalign*}
	b) \;\;\; \bigg(\frac{\partial\bar h}{\partial Y}\bigg)^2
	& = 
	\frac{3}{2}\, q_c^2 B_1^* C_1^*\epsilon^{i \,q_c x} + ... + c.c. & \; .
	\end{flalign*}
	
	Since the same adiabatic elimination from the critical ripples applies from the critical
	ripples applies for $\partial A_2 / \partial \tau $, the amplitude equation for $A_1$
	becomes:

	\begin{eqnarray*}
	\frac{\partial A_1}{\partial \tau} 
	&=& 
	\epsilon A_1 +\nu_A B_1^* C_1^* - u_A |A_1|^2 A_1 + ...
	\end{eqnarray*}

	\noindent where:
	
	\begin{eqnarray}
	\label{eq:ampA1}
	\nu_A = \frac{q_c^2}{2}\, ( 3\, \bar\nu_y - \bar\nu_x ) ,\;\; \textrm{and} 
	\;\;u_A  \approx \frac{4\, \bar\nu_x^2}{9\, \bar K} \; .
	\end{eqnarray}

	Then, the amplitude equation for $B_1$ is obtained through the same procedure. The
	equations for the first two harmonics of this second mode are  :
	
	\begin{eqnarray*}
	\frac{\partial B_1}{\partial \tau} &=& (\epsilon - \Omega) B_1 + \nu_B A_1^* C_1^*
	+ q_c^ 2 (\bar\nu_x + 3\, \bar\nu_y) B_1^* B_2 + ...
	\\[3mm]
	\frac{\partial B_2}{\partial \tau} &=& (\epsilon - 4\, \Omega - 9\, \bar K q_c^4) B_2
	- \frac{q_c^2 \bar\nu_x + 3 q_c^2\, \bar\nu_y}{4} B_1^2 + ...
	\end{eqnarray*}

	\noindent with $\Omega = \frac{3}{4} (|\mu|-|\nu|) q_c^2 $ and $\nu_B = q_c^2 \bar\nu_x$. Once again, adiabatic 
	elimination allows us to obtain the following relation between amplitudes:
	
	\begin{eqnarray*}
	B_2 &\approx& \frac{q_c^2 (\bar\nu_x + 3\, \bar\nu_y)}
	{4\,(\epsilon - 4\, \Omega - 9\, \bar K q_c^4)} B_1^2 \; .
	\end{eqnarray*}

	Hence, the amplitude equation for $B_1$ is:

	\begin{eqnarray}
	\label{eq:ampB1}
	\frac{\partial B_1}{\partial \tau} 
	&=& 
	(\epsilon - \Omega) B_1 +\nu_B A_1^* C_1^* - u_B |B_1|^2 B_1 + ...
	\end{eqnarray}

	\begin{eqnarray*}
	\textrm{with} \;\;\; u_B \approx \frac{q_c^4 (\bar\nu_x + 3\, \bar\nu_y)^2}{4\, ( 9\, \bar K q_c^4 + 4\, \Omega)}
	\;\;\; \textrm{for small } \epsilon \; .
	\end{eqnarray*}

	Similarly, the amplitude equation for $C_1$ is:

	\begin{eqnarray}
	\label{eq:ampC1}
	\frac{\partial C_1}{\partial \tau} 
	&=& 
	(\epsilon - \Omega) C_1 +\nu_C B_1^* A_1^* - u_C |C_1|^2 C_1 + ...
	\end{eqnarray}

	\noindent where $\nu_C = \nu_B$ and $u_C = u_B$. Therefore, the resulting hexagonal patterns
	present a structure of the type:
	
	\begin{eqnarray*}
	h = A \,\textrm{cos}(q_c\, x) + 2\,B \,\textrm{cos} \bigg(\frac{1}{2}\,q_c\, x \bigg)
	 \,\textrm{cos} \bigg(\frac{\sqrt{3}}{2}\, q_c\, y \bigg) \; .
	\end{eqnarray*}

	In case the damping $\alpha$ is too low, Eqs. \ref{eq:ampA1}-\ref{eq:ampB1}-\ref{eq:ampC1} do not
	hold to describe the resulting behavior, since an increasing number of harmonics may become
	unstable and more couples between smaller and larger scales will occur. However, if the
	damping is sufficiently high, the hexagonal structures should stabilize according to this
	this weakly nonlinear analysis. For the numerical results that will be presented in the 
	following sections, we were able to observe such hexagonal patterns for late time, based on 
	the previously described triad of modes $\vec{q}_1$, $\vec{q}_2$ ,and $\vec{q}_3$.



\section{Code Verification}
\label{sec:verification}

	In consonance with the guidelines for the manufactured solution 
	construction by Roache \cite{roache2002code}, an artificial solution was developed 
	considering spatio-temporal variations of a surface:
	
	\begin{eqnarray}
	\bar{h}_{m} &=& h_{o} + h_{xy} \, \mathrm{sin}\left(\dfrac{a_{x} \pi x}{L}\right) \,
	                                 \mathrm{cos}\left(\dfrac{a_{y} \pi y}{L}\right) \, \mathrm{e}^{\;bt} \; .
	\end{eqnarray}
	
	The parameter values employed in the manufactured solution and in the differential 
	equation are presented on Table \ref{table:mmsparameters}. Although the artificial solution
	does not need to be realistic for code verification, the chosen values were coherent with
	the studied simulations. Besides, the manufactured solution presents a periodic nature, which 
	is also the case of the posed problem. All tests were run in a quare domain with L = 256, 
	and a time step of $\Delta \tau = 0.1$.

	\begin{table*}[!htbp]
	\centering
	\ars{1.2}
	\caption{Parameter values for the MMS}
	\begin{tabular}{@{}c c c c c c c@{}} \toprule [{1.5pt}]
	\textbf{MS Parameters}  && \textbf{Value} && \textbf{KS Parameters} && \textbf{Value} \\ 	
	\midrule
	
		$h_{o}$    &&      0         &&    $\bar{K}$    &&   5        \\
		$h_{xy}$   &&   $\;10^{-4}$   &&    $a_{\eta}$    &&    4       \\
		$a_{x}$	   &&      2         &&    $\theta$     &&    \SI{30}{\degree}  \\
		$a_{y}$	 &&      2           &&    $\alpha$     &&   0.15       \\
		$b$	          &&   -1/200             &&        &&              \\
	
 	\bottomrule [{1.5pt}]

	\end{tabular}
	\label{table:mmsparameters}
	\end{table*}

\begin{figure}[ht!]
		\centering
		\includegraphics[width=0.6\textwidth]{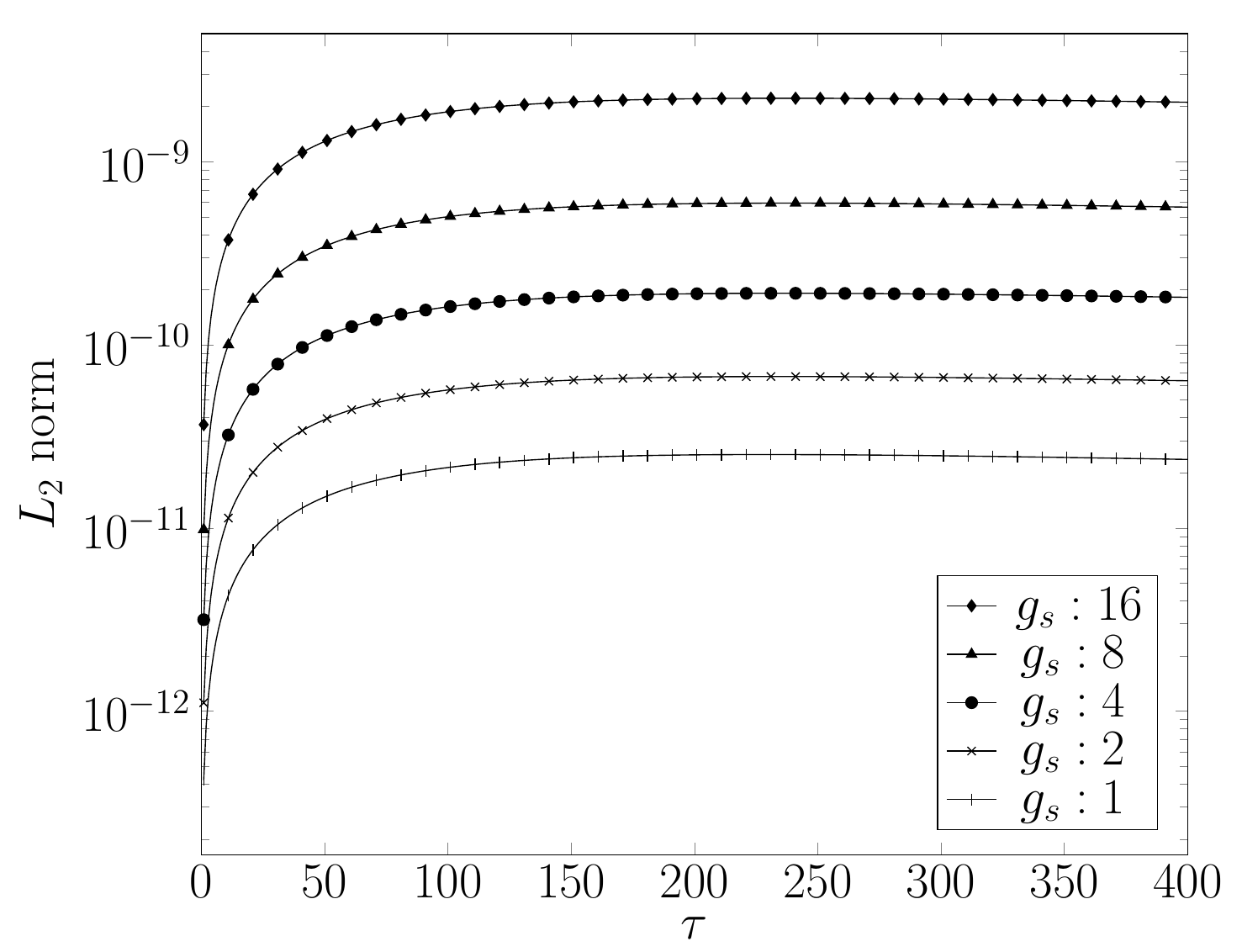}
		\caption{Evolution in time of the $L_{2}$ norm comparing different grid spacing $g_s$.}
		\label{fig:c4l2}
	\end{figure}

\begin{figure}[ht!]
		\centering
		\includegraphics[width=0.6\textwidth]{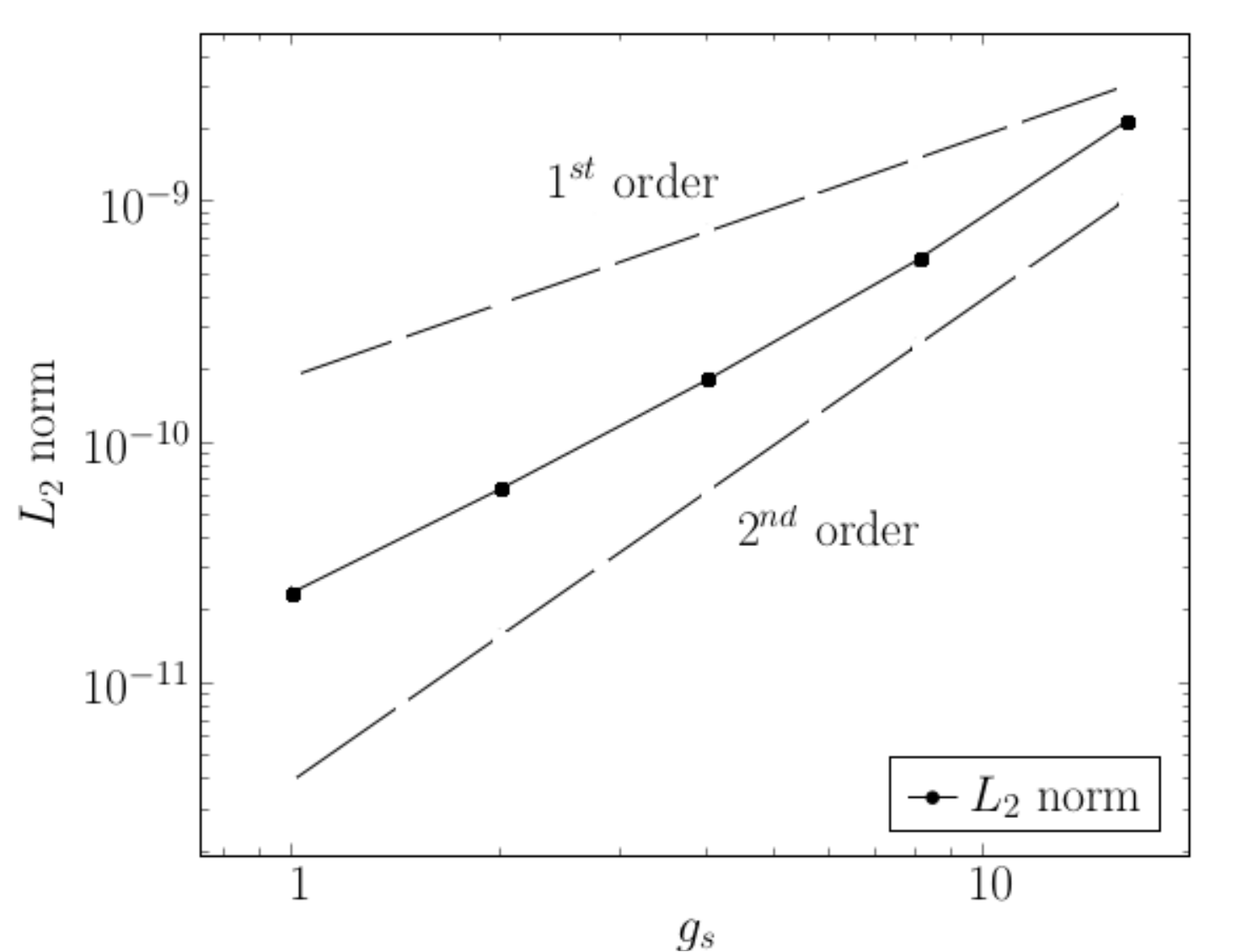}
		\caption{$L_{2}$ norm of the surface height for the manufactured solution. First 
		and second order error lines are also displayed for comparison.}
		\label{fig:c4l2order}
	\end{figure}

	Figures \ref{fig:c4l2} and \ref{fig:c4l2order} display  how the spatial
    grid-refinement affects the global discretization error. The error tends toward
    second-order convergence with respect to grid spacing $g_s$ for coarser meshes.
    However, as the grid gets more refined (smaller $g_s$), the error stands between
    first and second-order convergence.
	
	The manufactured solution was unstable for the analyzed differential equation, which grows
	rapidly in time until its saturation, shadowing the contribution of the source term. For
	this reason, we limited the range of analysis to a stable region for the artificial
	function, where the numerical solution converges to the analytical solution. As we will
	later discuss, this region corresponds to a stage before the emergence of hexagonal modes.


\section{Scheme Stability}
\label{sec:stability}	
	
	One important issue which concerns the simulations is the time and mesh size 
	selection, such that we seek reasonable choices inside the stable region of the present 
	semi-implicit scheme. Here, we study the time step and grid spacing variation effect 
	regarding the pattern evolution by tracking the $L_{1}$ norm evolution in time. 
	During the simulations, we monitored the pattern's evolution by the $L_1$ 
	norm rate of change in time, which indicates how fast the structure is changing between 
	the current and previous time step, normalized by the spatial average of the surface height 
	absolute value. This norm rate of variation is denoted as:

	\begin{eqnarray}
		L_{1,t} &=& 
		\frac{1}{\raisebox{-0.75ex}{$\Delta \tau$}}
		\frac{\sum_{\;ij}\,|\bar h^{\;n+1}_{\;ij}-\bar h^{\;n}_{\;ij}|}
		{\raisebox{-0.75ex}{$\sum_{\;ij}\,|\bar h^{\;n+1}_{\;ij}|$}} \; .
	\end{eqnarray}
	
	The computational effort was also measured, being related to the number of internal 
	iterations when comparing results for a same grid spacing. Two different grids were
	used for the stability tests, as follows:
	
	\begin{enumerate}
	\item Case A: $\Delta X$ = $\Delta Y$ = 2, 64$\times$64 points in a domain 128$\times$128
	\item Case B: $\Delta X$ = $\Delta Y$ = 1, 128$\times$128 points in a domain 128$\times$128
	\end{enumerate}	

	Both cases start with a monomodal initial pattern $q_{0}\vec{1}_{x}$, presenting four 
	wavelengths in the domain. The critical wavelength (related to the critical wavenumber 
	$q_{c}$) is approximately 18: each of them would be represented by 9 points for Case A and 
	by 18 points for Case B. The parameters adopted for the tests are displayed on Tab.
	\ref{table:parameters}.
	
	\begin{table*}[!htbp]
	\caption{Parameters value and description}
	\centering
	\ars{1.2}
	\begin{tabular}{@{}l c c c l @{}} \toprule [{1.5pt}]
	
	\label{table:parameters}
	
	Parameter && Value && Description \\ \midrule
	
	$\alpha$ && 0.15 && damping coefficient\\
	$\bar{K}$ && 5 && surface diffusion effects\\
	$\theta$ && \SI{30}{\degree} && beam's angle of incidence\\
	$a_{\eta}$ && 4 && penetration depth/width of energy distribution\\
	 \bottomrule [{1.5pt}]
		
	\end{tabular}
	\end{table*}

	We assumed a pattern as stationary if the criterion $L_1 < 10^{-7}$ was reached for the 
	temporal evolution. On the other hand, the simulation would be stopped if the $L_1$ curve
	 demonstrated clearly a behavior converging to a fixed value (or oscillating around it). 
	 Even though $L_{1,t}$ is a rate of change and not the $L_1$ norm, we will
	refer $L_{1,t}$ just as $L_1$ for the sake of simplicity.

\begin{figure*}[ht!]
    	\centering
   	\begin{subfigure}[t]{0.6\textwidth}
   		\centering
       		\includegraphics[width=0.95\linewidth]
        		{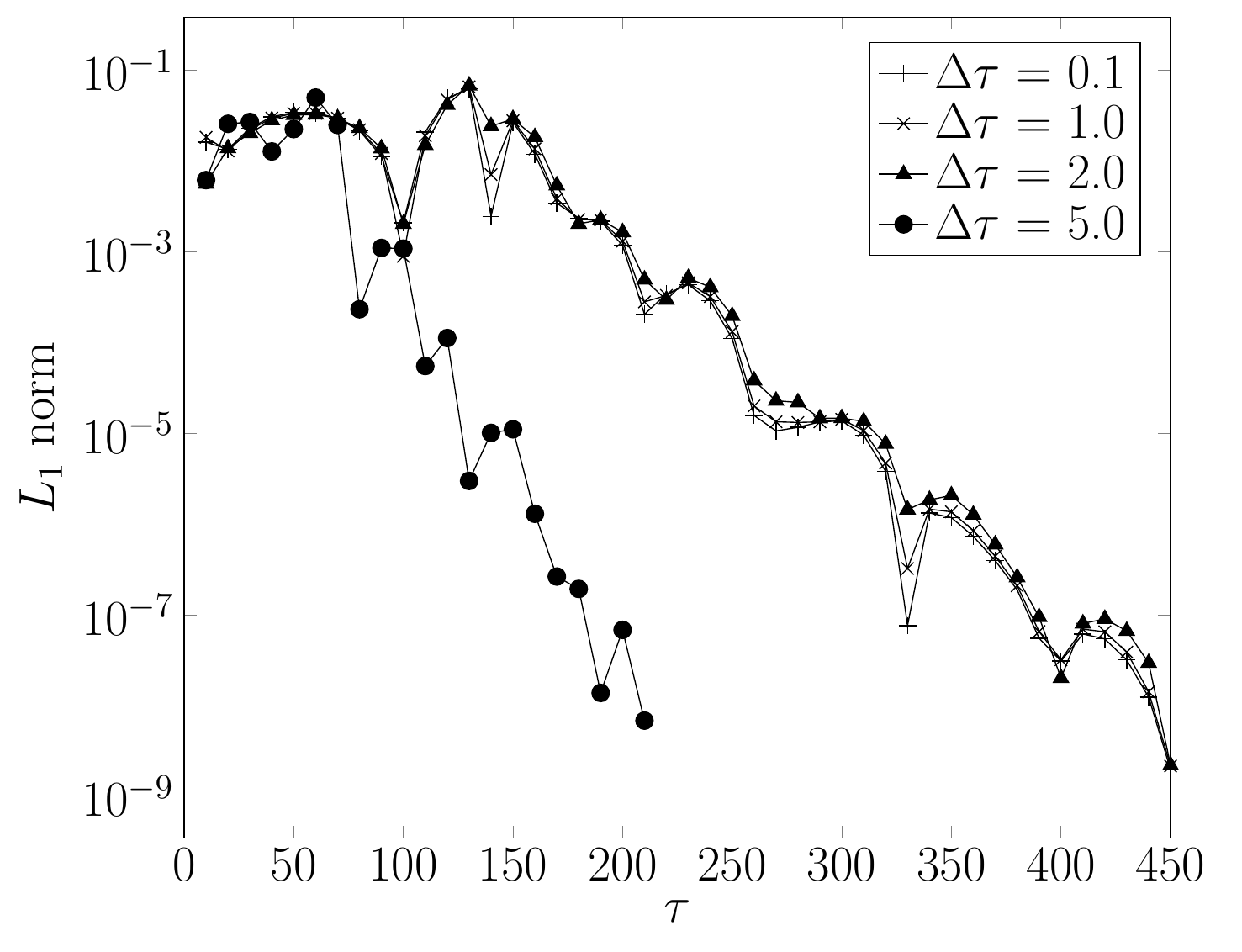}
       		\caption{Case A: $\Delta X = 2$, 64$\times$64 nodes in a domain 128$\times$128}
       		\label{fig:stabA}
    	\end{subfigure}
    	\begin{subfigure}[t]{0.6\textwidth}
        		\centering
        		\includegraphics[width=0.95\linewidth]
        		{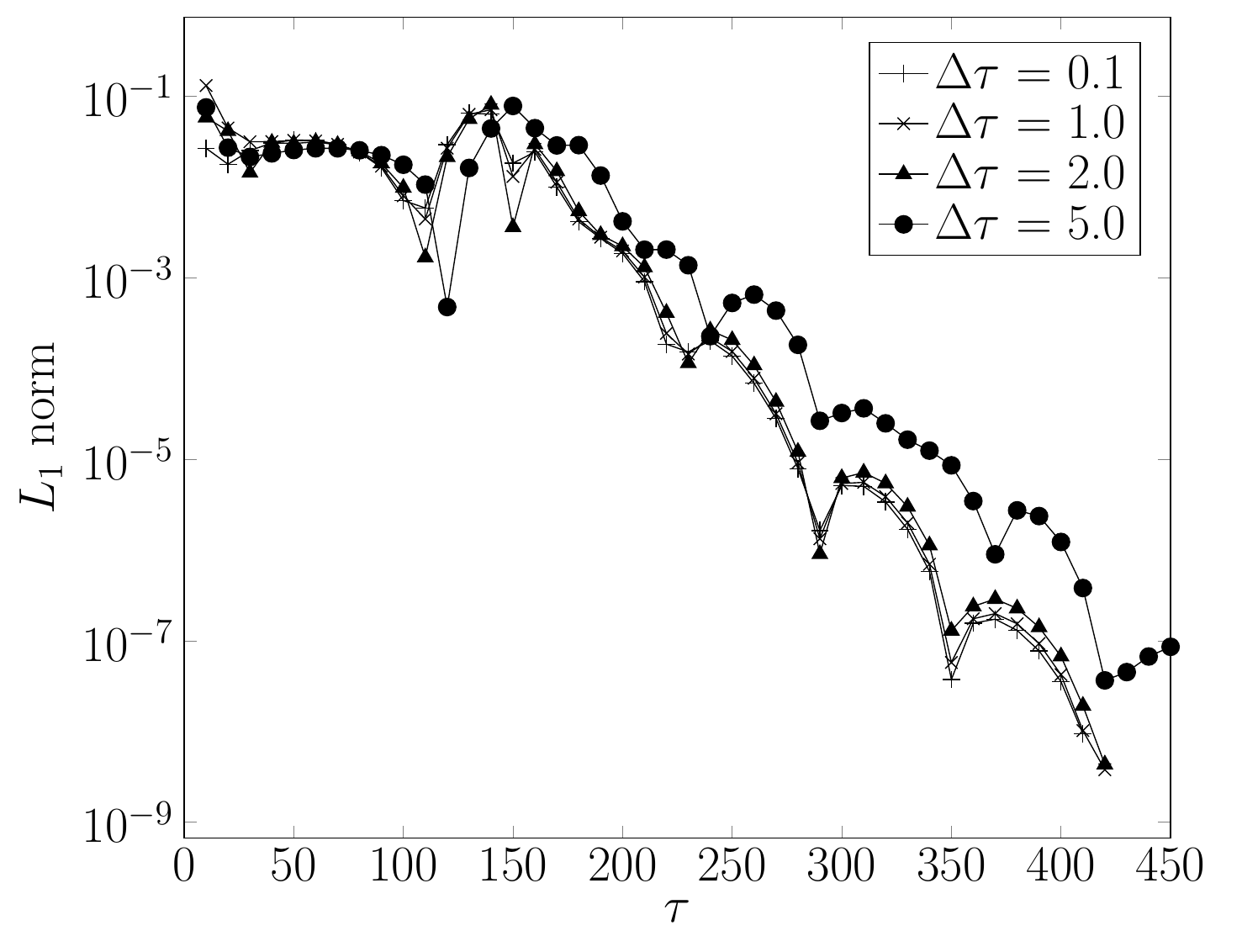}
        		\caption{Case B: $\Delta X = 1$, 128$\times$128 nodes in a domain 
        		128$\times$128}
        		\label{fig:stabB}
    	\end{subfigure}
    	\vspace{2mm}
    	\caption{$L_{1}$ norm as a measure for stability regarding time step.}
    	\label{fig:stab}
	\end{figure*}

	Figure \ref{fig:stab} compares the structure evolution through the $L_1$ norm to observe
	when the results would diverge for an increasing time step. 

	From Fig. \ref{fig:stabA} (Case A) we observe slight deviations in the $L_1$ norm 
	evolution for $\Delta \tau = 2.0$ when comparing to inferior time steps values,
    while $\Delta \tau = 5.0$ diverges completely from the others. Regarding Fig.
    \ref{fig:stabB} (Case B), the divergence for $\Delta \tau = 5.0$ is also clear,
    but it's more coherent with the smaller time steps than Case A, as expected from
    a more refined mesh. This time, $\Delta \tau = 2.0$ is more consistent with the
    smaller ones, and would be accepted for the simulations. Even so, we decided to
    operate with up to $\Delta \tau = 1.0$ for both $\Delta X = 1$ and 2, which is a
    more conservative approach. Nevertheless, represents a significant increase in
    time step (inside the stable domain) when compared to some of the aforementioned
    works that adopted explicit schemes
    \cite{paniconi1997stationary,facsko2004dissipative}.


\section{\textbf{Anisotropic DKS evolution on preexisting patterns}}
\label{sec:dks}

	Results regarding the evolution of the anisotropic DKS on preexisting patterns 
	will be discussed next. The simulation results were obtained with the parameters 
	$\bar{K}$, $\theta$, $a_\eta$ and $\alpha$ as shown in Tab. 2, using a mesh
    consisting of 256$\times$256 points, and a dimensionless grid spacing of
    $\Delta X = 2$. The reason for this choice is that $K$ is givern by:

    \begin{equation*}
      K \; = \; \frac{D_s \, \rho \, n_d}{N^2 \, k_B \, T}
      {\rm exp} \left(-\frac{\Delta E}{k_B \, T} \right)
    \end{equation*}

    \noindent where $D_s$ is the surface self-diffusivity, $\rho$ is the surface free
    energy per unit area, $n_d$ the density of diffusing atoms, $N$ the atomic
    density, $k_B$ the Boltzmann constant, $\Delta E$ the activation energy and
    $T$ the temperature. We are particularly interested in high temperatures
    ($\bar{K} > 1$), such that diffusion is enhanced, acting on similar scales
    with sputtering terms. Then, by setting $T = 500$ K, we choose a realistic
    activation energy of $\Delta E = 1.25$ eV (typically in the range of
    1 - 2.6 eV for small clusters of atoms), and do the same for the remaining
    parameters. This leads to the dimensionless $\bar{K} = 5$, and to the choice
    of $a_\eta = 4$. Specific experimental systems can be studied by adjusting
    each of these parameters, depending for instance on the target material,
    and on the flux and energy of incoming ions. Those looking for further
    experimental data to perform simulations should visit some of aforementioned
    references \cite{makeev2002morphology,keller2010ion,valbusa2002nanostructuring}.

	\subsection{[Case 1] Initial pattern with $\vec{q} = q_{0}\vec{1}_{X}$, 2 wavelengths}

\begin{figure*}[htbp!]
   	\begin{subfigure}[t]{0.48\textwidth}
       		\includegraphics[width=1.0\linewidth]
        		{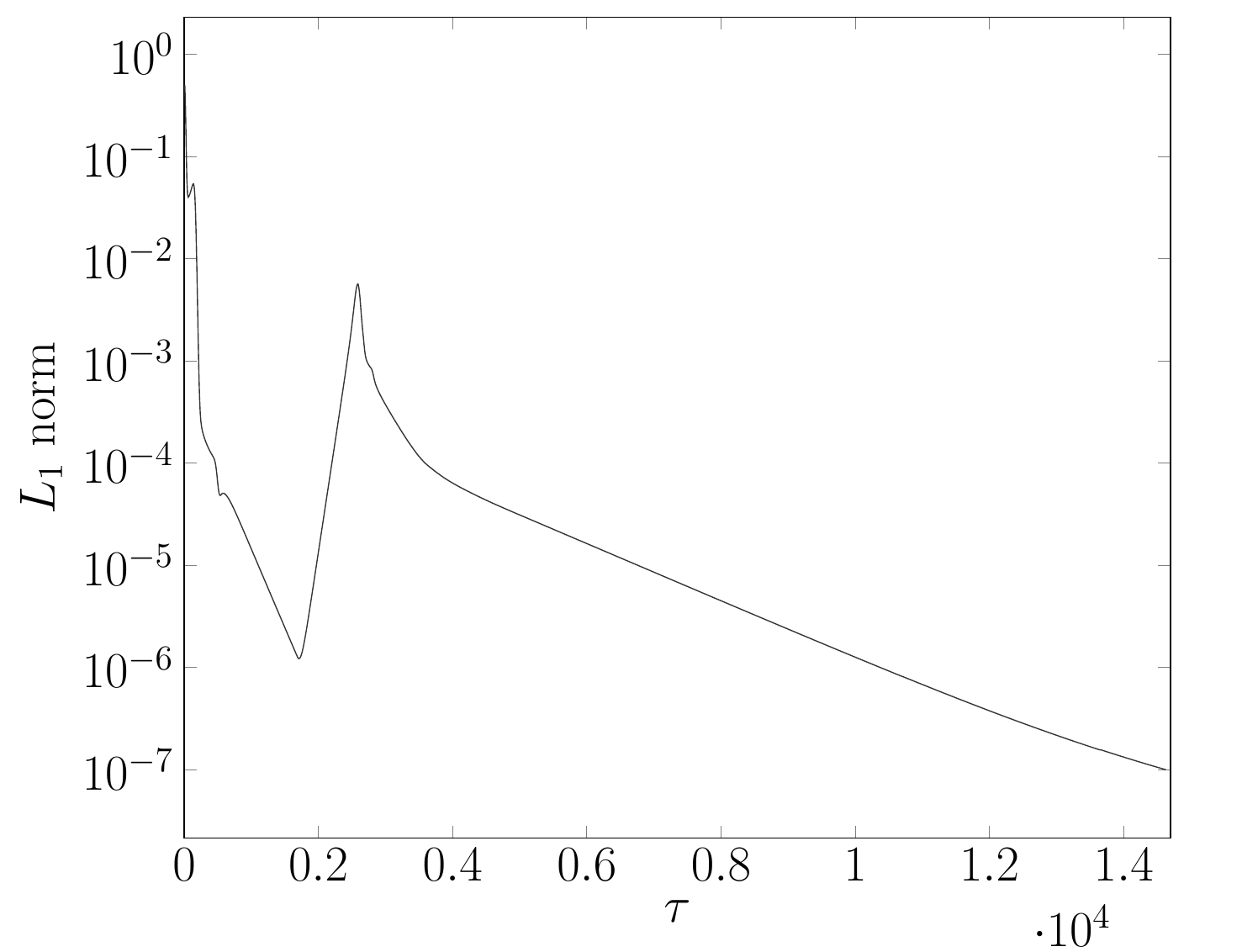}
       		\caption{$L_{1}$ norm rate of variation}
    	\end{subfigure}
    	\hspace{4mm}
   	\begin{subfigure}[t]{0.48\textwidth}
       		\includegraphics[width=1.0\linewidth]
        		{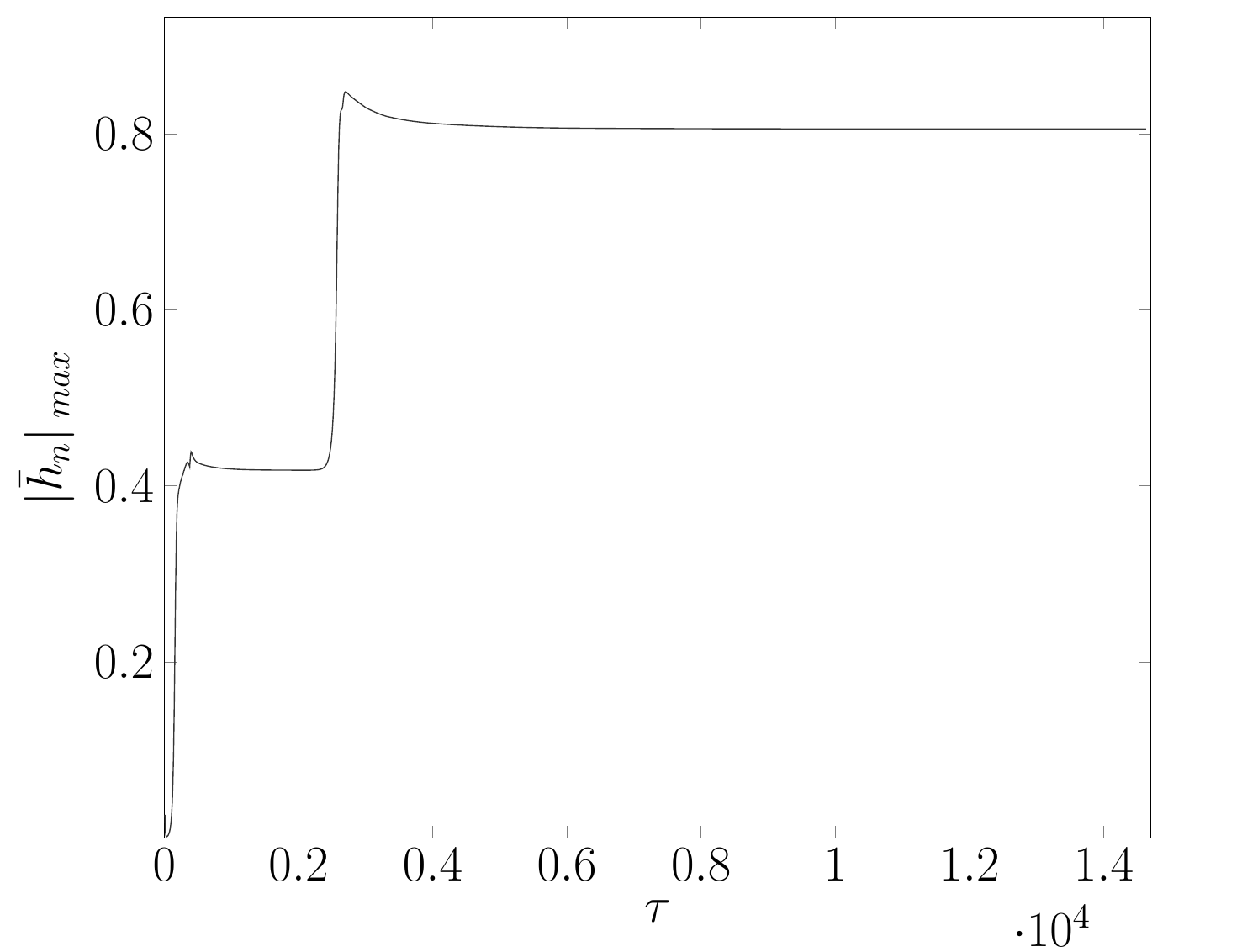}
       		\caption{Maximum $\abs{\bar{h}^{\;n}}$}
    	\end{subfigure}
    	\caption{$L_{1}$ norm rate of variation, and maximum values for $\bar{h}^{\;n}$ 
    	as a function of time $\tau$, for an initial pattern with $\vec{q} = q_{0}\vec{1}_{X}$ (inside
    	the stable domain), $\Delta\tau = 0.1$, $\Delta X = \Delta Y = 2$, on a 256 $\times$ 256
    	nodes mesh. Beam angle is set at \SI{30}{\degree}.}
    	\label{fig:case1info}
	\end{figure*}

	We adopt an initial pattern consisting of an 1D structure with a wavenumber $q_0 = 2.4544 
	\cdot 10^{-2}$ in the $\vec{1}_{X}$ direction (2 starting wavelengths in the system), situated 
	inside the stable domain. Figure \ref{fig:case1info} presents the time evolution of the $L_1$ 
	norm, internal iterations, and maximum height $\abs{\bar{h}^{\;n}}$ for this case, with a zoom 
	into the initial regimes to display how the structure diminishes and rises. 
	
\begin{figure*}[h!]
    	\centering
   	\begin{subfigure}[t]{0.47\textwidth}
       		\includegraphics[width=1.20\linewidth]
        		{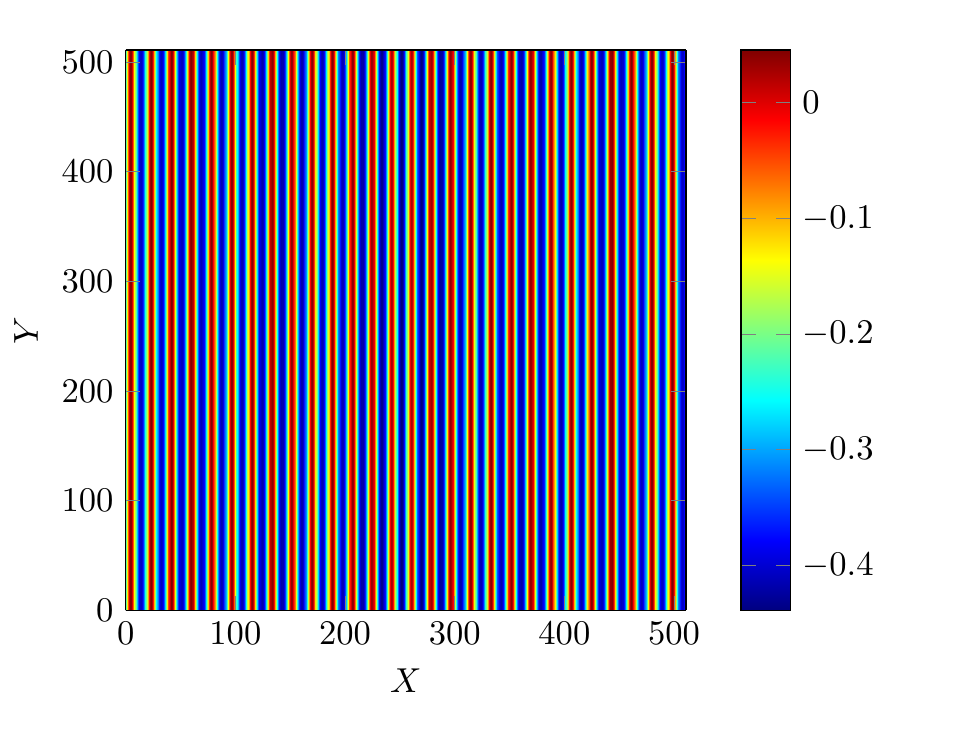}
    	\end{subfigure}
    	\hspace{5mm}
    	\begin{subfigure}[t]{0.47\textwidth}
        		\centering
        		\vspace{-54mm}
        		\includegraphics[width=0.95\linewidth]
        		{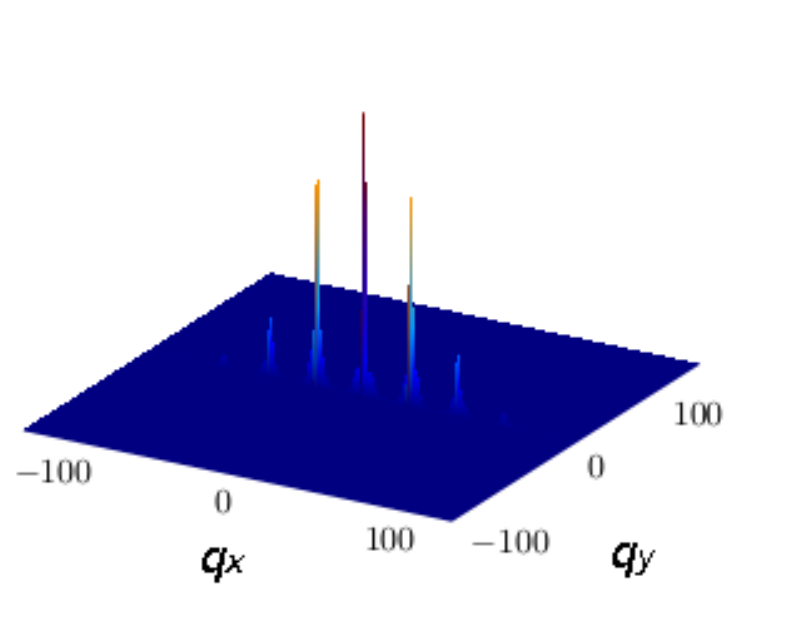}
    	\end{subfigure}
    	   	\begin{subfigure}[t]{0.47\textwidth}
       		\includegraphics[width=1.20\linewidth]
        		{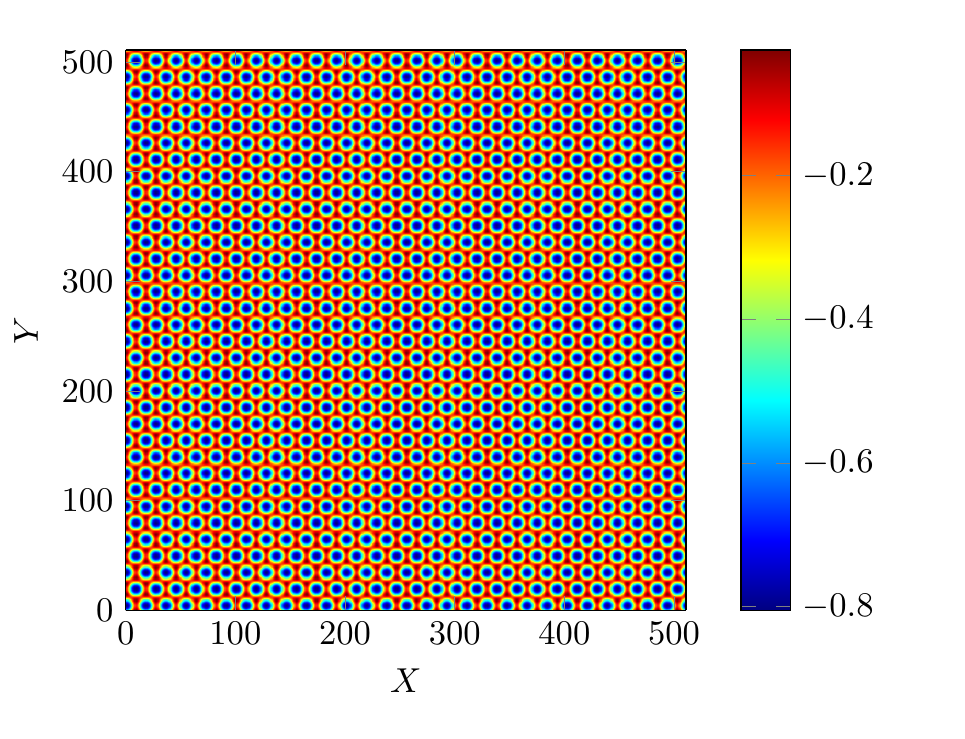}
    	\end{subfigure}
    	\hspace{5mm}
    	\begin{subfigure}[t]{0.47\textwidth}
        		\centering
        		\vspace{-54mm}
        		\includegraphics[width=0.95\linewidth]
        		{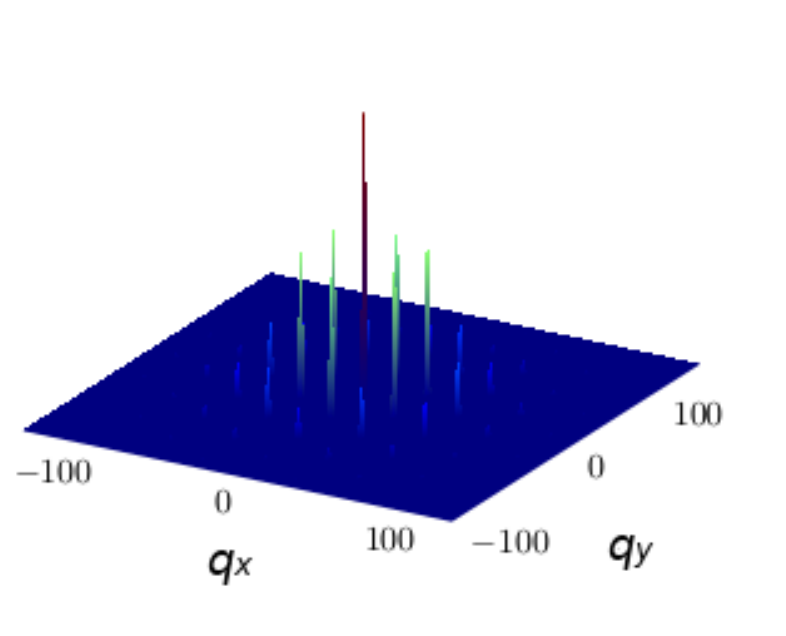}
    	\end{subfigure}
    	\caption{Surface height values $\bar{h}^{\;n}$ and their respective Fourier
            Transform for $\tau = 400$ (upper panels) and $\tau = 14,630$, with an
            initial monomodal pattern presenting $\vec{q} = q_c\vec{1}_X$ and the
            beam angle set at \SI{30}{\degree}. The 1D structure is destabilized
            due to the emergence of the hexagonal modes.}
    	\label{fig:case1}
	\end{figure*}

	From $\tau = 0$ to $\tau = 400$, it is possible to follow the nanostructuration of
	an 1D pattern  with a wavenumber near $q_c$ (upper panels of Fig. \ref{fig:case1}). Based 
	on the $L_1$ curve and on the maximum absolute height value, one could think that a steady 
	state was already reached about $\tau = 2,000$, since $\bar{h}^{\;n}$ stabilized in a minimum 
	of -0.42, and the pattern calmed down (see the continuous and pronounced fall of $L_1$). 
	Nonetheless, after $\tau = 2,000$, the hexagonal modes emerge, and the 1D structure 
	destabilizes. The $L_1$ curve rapidly ascends to its peak and then falls down, as the 
	structure reorganizes and stabilizes with the new hexagonal pattern. A defectless nanohole 
	pattern is attained for the steady state, with a minimum height frozen at -0.8, as seen in the 
	lower panels of Fig. \ref{fig:case1}. The system is very sensitive to the initial wavenumber,
	and slight variations of $q_c$ may lead to hexagonal patterns with remaining defects.

re
	\subsection{[Case 2] Initial pattern with $\vec{q} = q_{0}\vec{1}_{Y}$, 2 wavelengths}

\begin{figure*}[htbp!]
   	\begin{subfigure}[t]{0.48\textwidth}
       		\includegraphics[width=1.0\linewidth]
        		{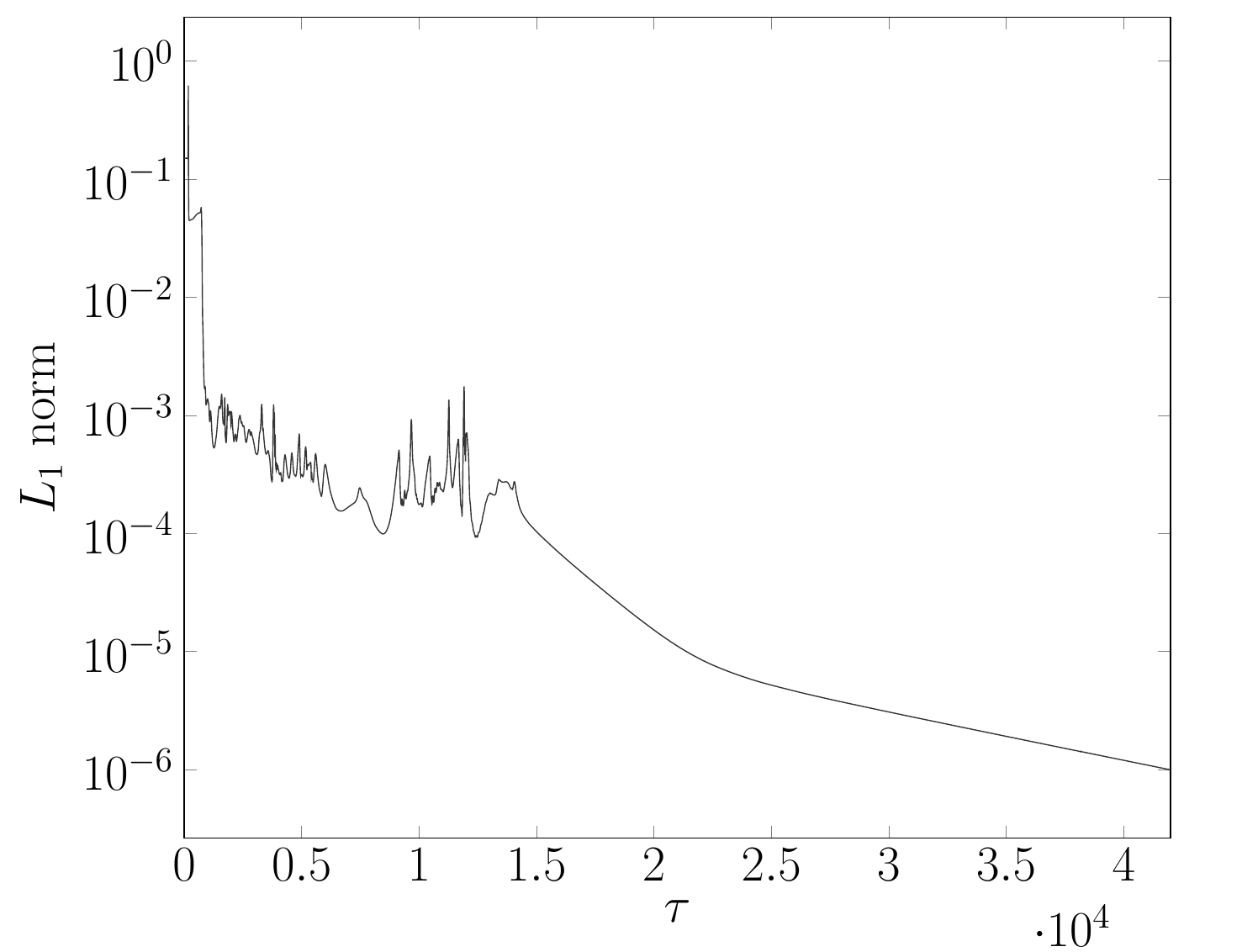}
       		\caption{$L_{1}$ norm rate of variation}
    	\end{subfigure}
    	\hspace{4mm}
   	\begin{subfigure}[t]{0.48\textwidth}
       		\includegraphics[width=1.0\linewidth]
        		{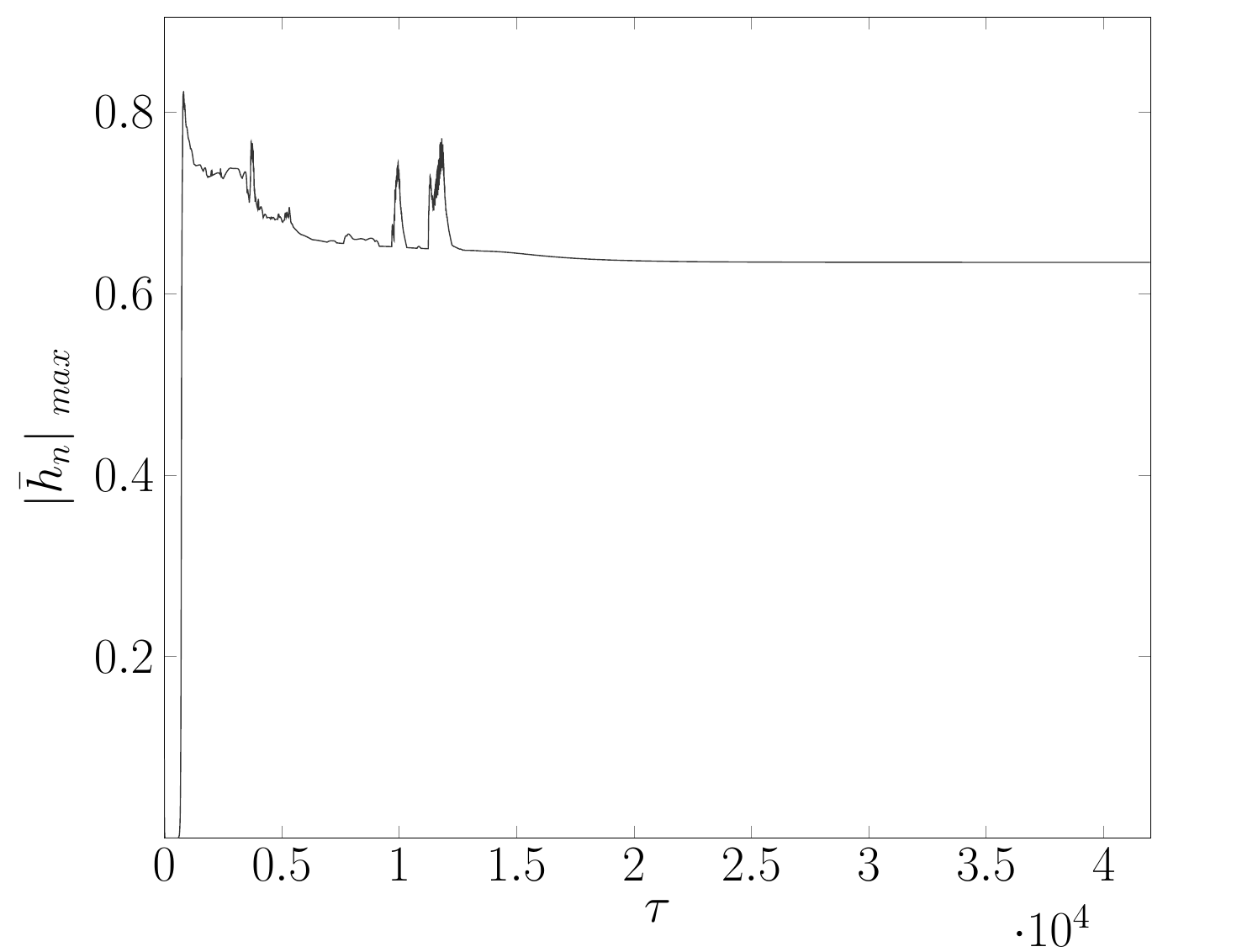}
       		\caption{Maximum $\abs{\bar{h}^{\;n}}$}
    	\end{subfigure}
    	\caption{$L_{1}$ norm rate of variation, and maximum values for
            $\bar{h}^{\;n}$ as a function of time $\tau$, for an initial pattern with
            $\vec{q} = q_{0}\vec{1}_{Y}$ (inside the stable domain),
            $\Delta\tau = 0.1$, $\Delta X = \Delta Y = 2$, on a 256 $\times$ 256
            nodes mesh. Beam angle is set at \SI{30}{\degree}.}
    	\label{fig:case2info}
	\end{figure*}

	The initial pattern for the second case is a monomodal $\vec{1}_Y$ surface,
	possessing the same number of wavelengths as Case 1, for the sake of comparison
	($q_0 = 2.4544 \cdot 10^{-2}$). This wavenumber also lies in the linearly stable
	domain, so we expect a similar behavior of the dynamics during the first stage.
	Figure \ref{fig:case2info} contains the time evolution of the $L_1$ norm, internal iterations,
	and maximum height $\abs{\bar{h}^{\;n}}$ for the second simulation. If we zoomed in the 
	initial regimes, it would be possible to see the 1D structure diminishing in height before the 
	emergence of hexagonal modes.
	
\begin{figure*}[h!]
    	\centering
   	\begin{subfigure}[t]{0.47\textwidth}
       		\includegraphics[width=1.20\linewidth]
        		{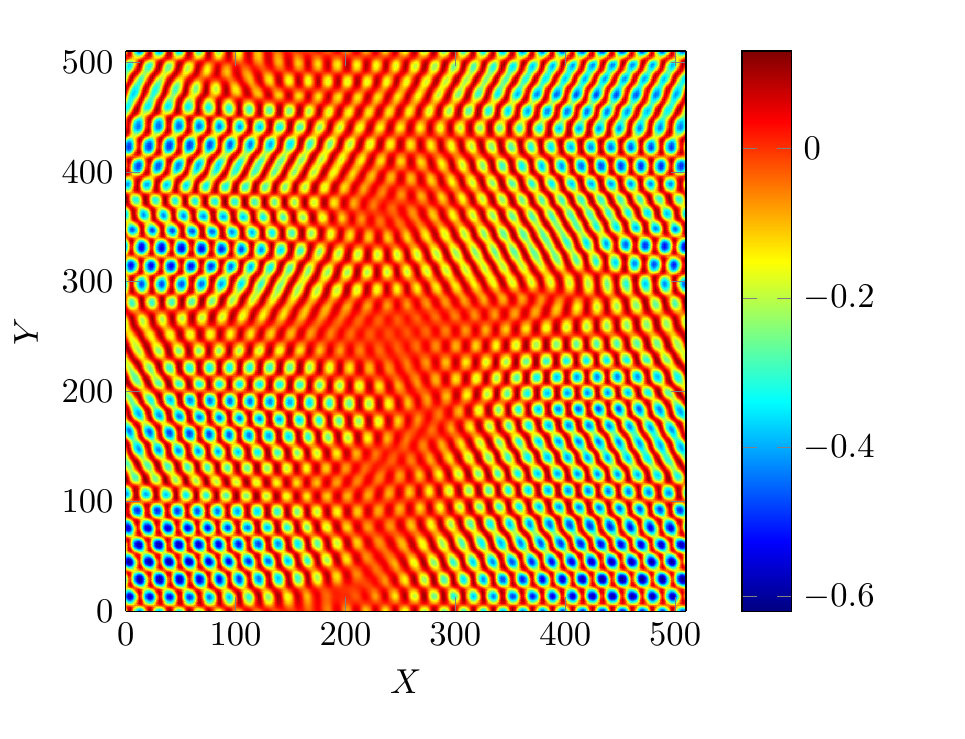}
    	\end{subfigure}
    	\hspace{5mm}
    	\begin{subfigure}[t]{0.47\textwidth}
        		\centering
        		\vspace{-54mm}
        		\includegraphics[width=0.95\linewidth]
        		{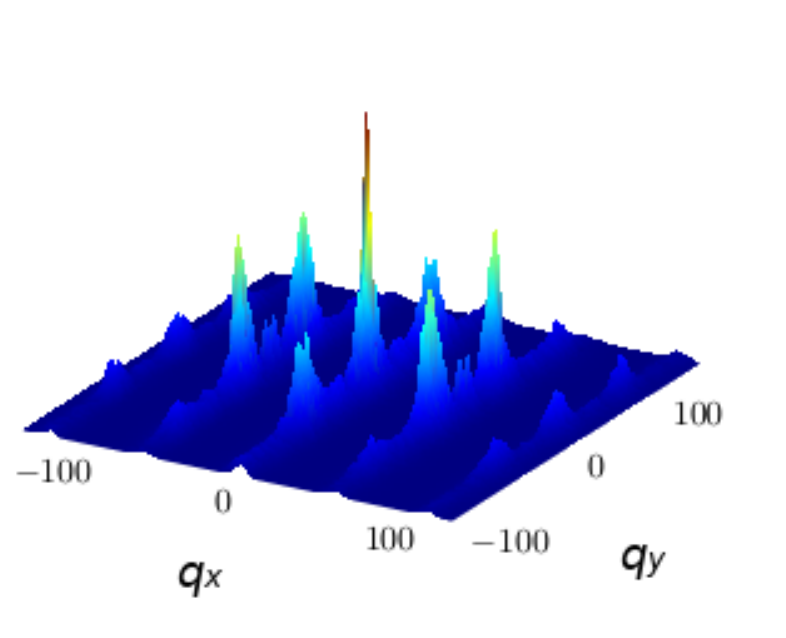}
    	\end{subfigure}
    	   	\begin{subfigure}[t]{0.47\textwidth}
       		\includegraphics[width=1.20\linewidth]
        		{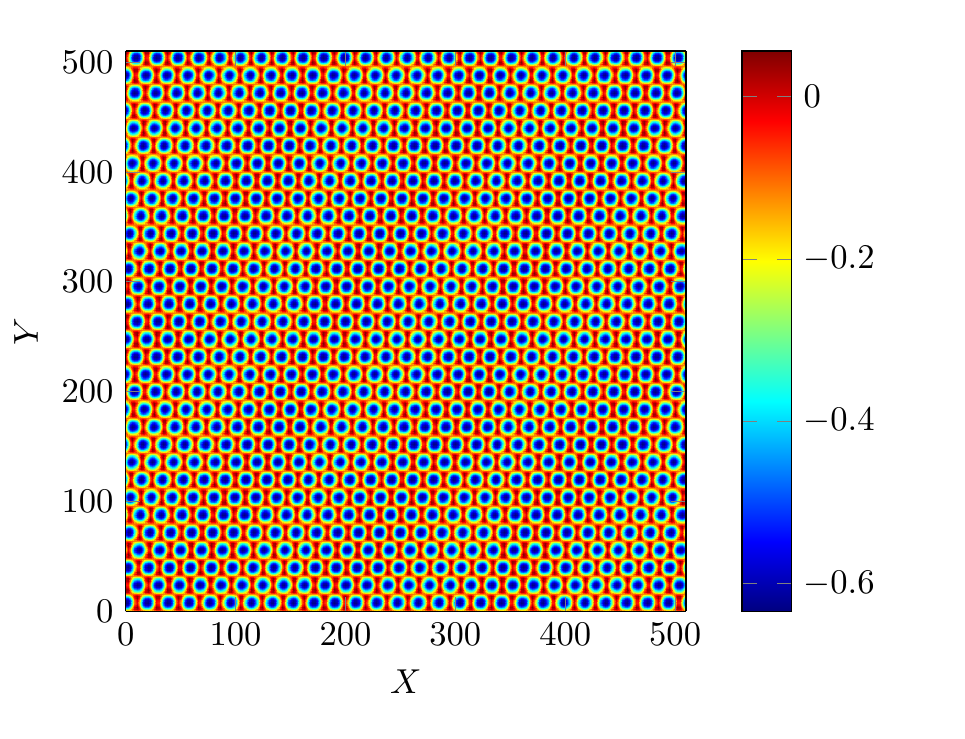}
    	\end{subfigure}
    	\hspace{5mm}
    	\begin{subfigure}[t]{0.47\textwidth}
        		\centering
        		\vspace{-54mm}
        		\includegraphics[width=0.95\linewidth]
        		{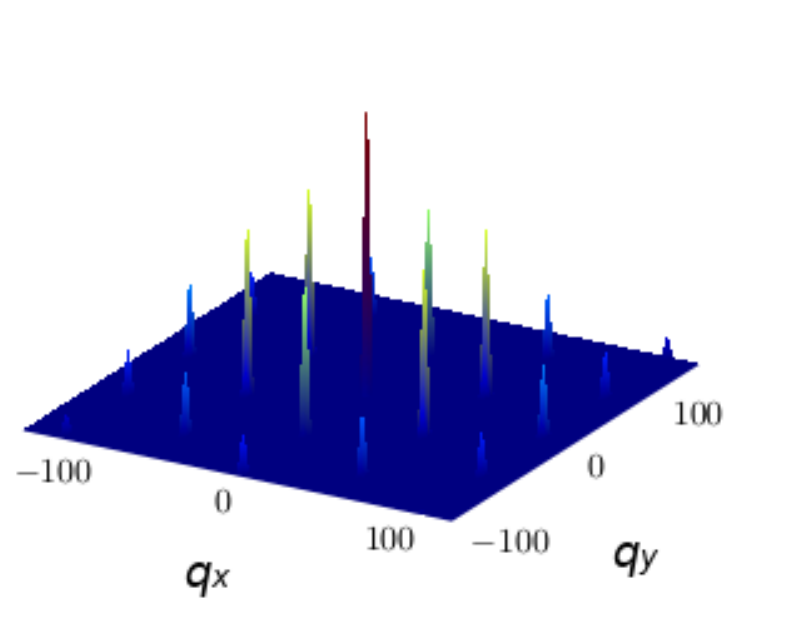}
    	\end{subfigure}
    	\caption{Surface height values $\bar{h}^{\;n}$ and their respective Fourier
            Transform for $\tau = 750$ (upper panels) and $\tau = 40,000$, with an
            initial monomodal pattern presenting $\vec{q} = q_c\vec{1}_Y$ and the
            beam angle set at \SI{30}{\degree}. After the structure diminishes
            (while in the stable domain), the nonlinearities lead to the emergence
            of hexagonal modes, quickly dominating the entire system.}
    	\label{fig:case2}
	\end{figure*}

	Case 2 does not display the formation of a 1D pattern with $\vec{q} \approx q_c\vec{1}_X$, as in 	
	Case 1. Between $\tau = 150$ and 200, the monomodal $\vec{1}_Y$ direction surface 
	disappears, and new ripples start to proliferate in the domain. Figure \ref{fig:case2} shows 
	that these ripples quickly lose their  orientation towards $\vec{1}_X$, and about $\tau = 750$ 	
	 we already have the formation of nanoholes. Therefore, the nanohole 
	structure takes over the system without an intermediate transition. 
	
	The steady state (lower panels of Fig. \ref{fig:case2}) consists of a well ordered defectless 
	nanohole pattern. Similarly to Case 1, this result is also physically consistent with the 
	sputtering phenomenon, as the height decay is expected from the removal of surface atoms, 
	alongside the rearrangement of the surface morphology. Besides, nanohole formation is one 
	of the organized patterns achieved by ion beam sputtering.
	
	\subsection{Damping tests}
	
	The linear damping term had an essential contribution to the acquisition of the nanohole
	pattern, as seen in the last section. According to the previous work of 
	Paniconi and Elder \cite{paniconi1997stationary}, three distinct solutions in the late time limit 
	might be expected for the DKS equation, depending on the parameter $\alpha$: periodic 
	large hexagonal morphology for higher values, an oscillatory or breathing hexagonal state for 
	middle values, and a spatiotemporal chaotic state for lower values. However, since the 
	present endeavor considers realistic coefficients related to the physics of sputtering, the 
	range of $\alpha$ values employed by Paniconi and Elder would not produce the same 
	effects. Besides, opposed to Paniconi and Elder, our study deals with an anisotropic DKS 
	equation. Thereafter, during the first moments, linear effects lead to the selection of a well 
	defined ripple direction; then, once nonlinear effects take over the system, cellular 
	structures will develop (clearly seen on the results for $\alpha = 0.15$).

\begin{figure*}[h!]
    	\centering
   	\begin{subfigure}[t]{0.48\textwidth}
   		\centering
       		\includegraphics[width=1.00\linewidth]
        		{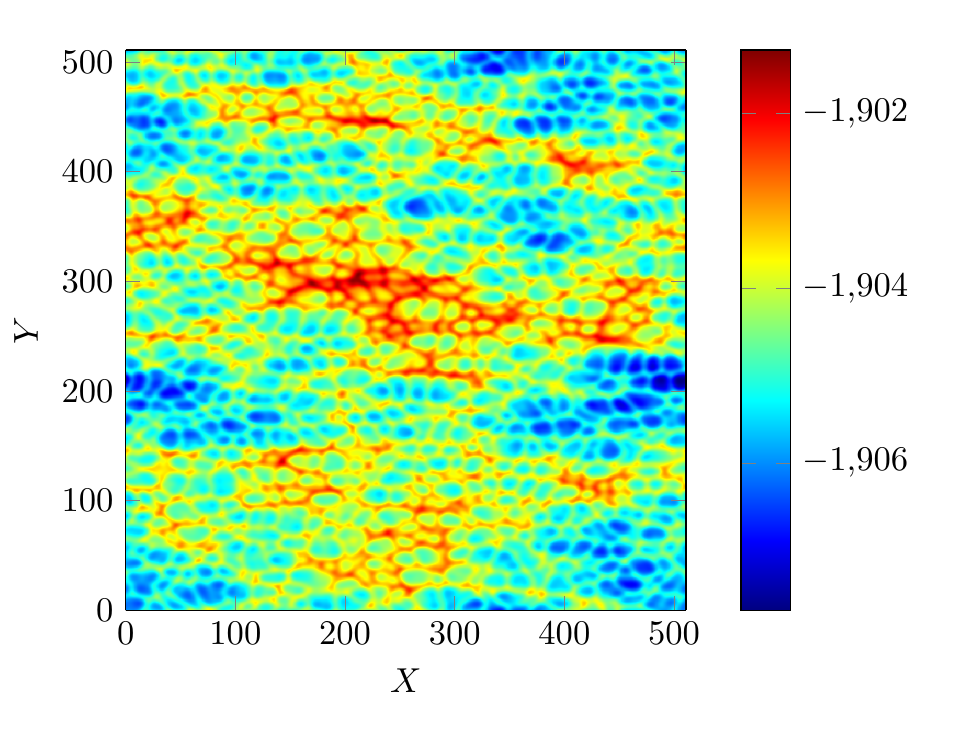}
       		\caption{Surface height $\bar{h}^n$ for $\tau = 11,803$}
       		\label{fig:chaos1a}
    	\end{subfigure}
    	\begin{subfigure}[t]{0.48\textwidth}
        		\centering
        		\includegraphics[width=0.95\linewidth]
        		{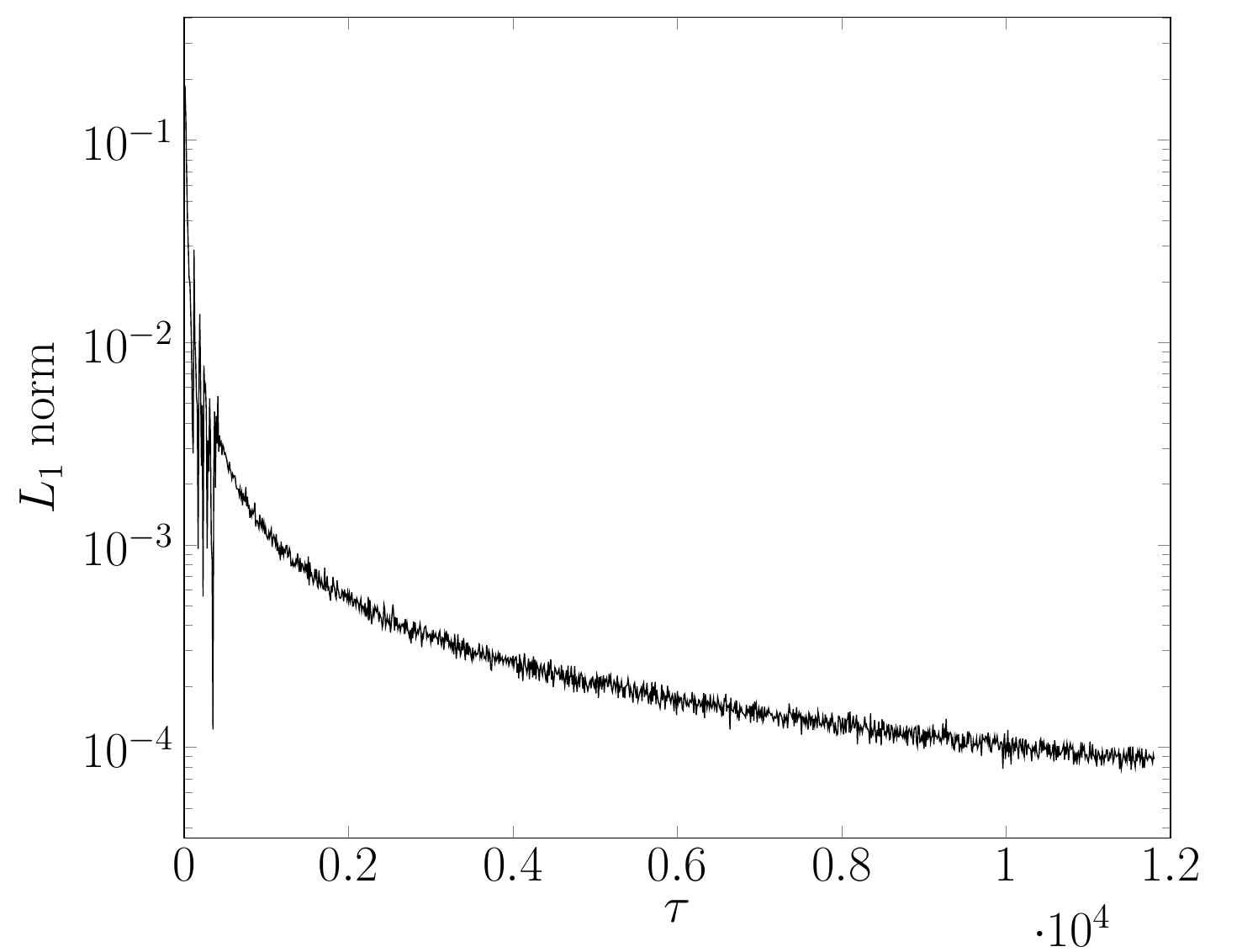}
        		\caption{$L_1$ norm evolution}
        		\label{fig:chaos1b}
    	\end{subfigure}
    	\vspace{2mm}
    	\caption{Numerical solution for a 2D anisotropic DKS equation -  Spatiotemporal chaotic 
    	pattern, with $\alpha = 0$ and $\theta = 30^\circ$.}
    	\label{fig:chaos1}
	\end{figure*}

	The undamped solution is shown in Fig. \ref{fig:chaos1}, for $\alpha = 0$. The initial
	condition presented a wavenumber $q_c = 1.7181 \cdot 10^{-1}$ (14 wavelengths in the
	system). A disordered chaotic cellular structure is obtained for late time, with large 
	variations of cell size and shape, as displayed in Fig. \ref{fig:chaos1a} for $\tau = 11,803$. 
	From the $L_1$ curve (Fig. \ref{fig:chaos1b}), we can see that the chaotic pattern is 
	reached within $\tau  = 500$. While a steady state isn't reached for the analyzed period, it's 
	clear that the evolution dynamics are much slower during late time. 

\begin{figure*}[h!]
    	\centering
   	\begin{subfigure}[t]{0.48\textwidth}
   		\centering
       		\includegraphics[width=1.00\linewidth]
        		{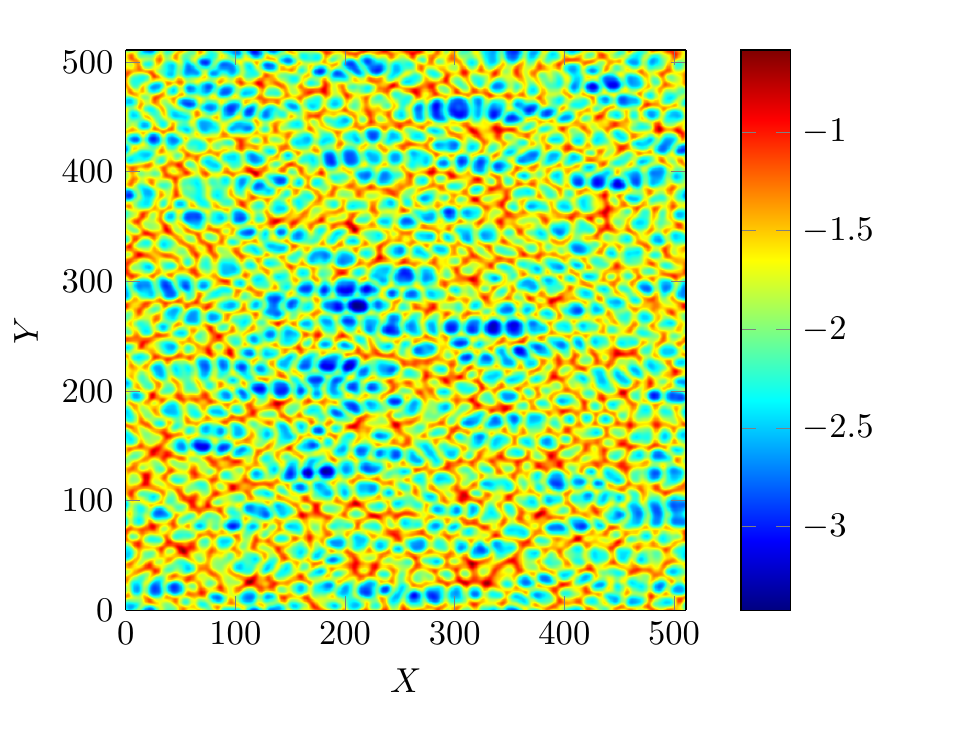}
       		\caption{Surface height $\bar{h}^n$ for $\tau = 11,750$}
       		\label{fig:chaos2a}
    	\end{subfigure}
    	\begin{subfigure}[t]{0.48\textwidth}
        		\centering
        		\includegraphics[width=0.95\linewidth]
        		{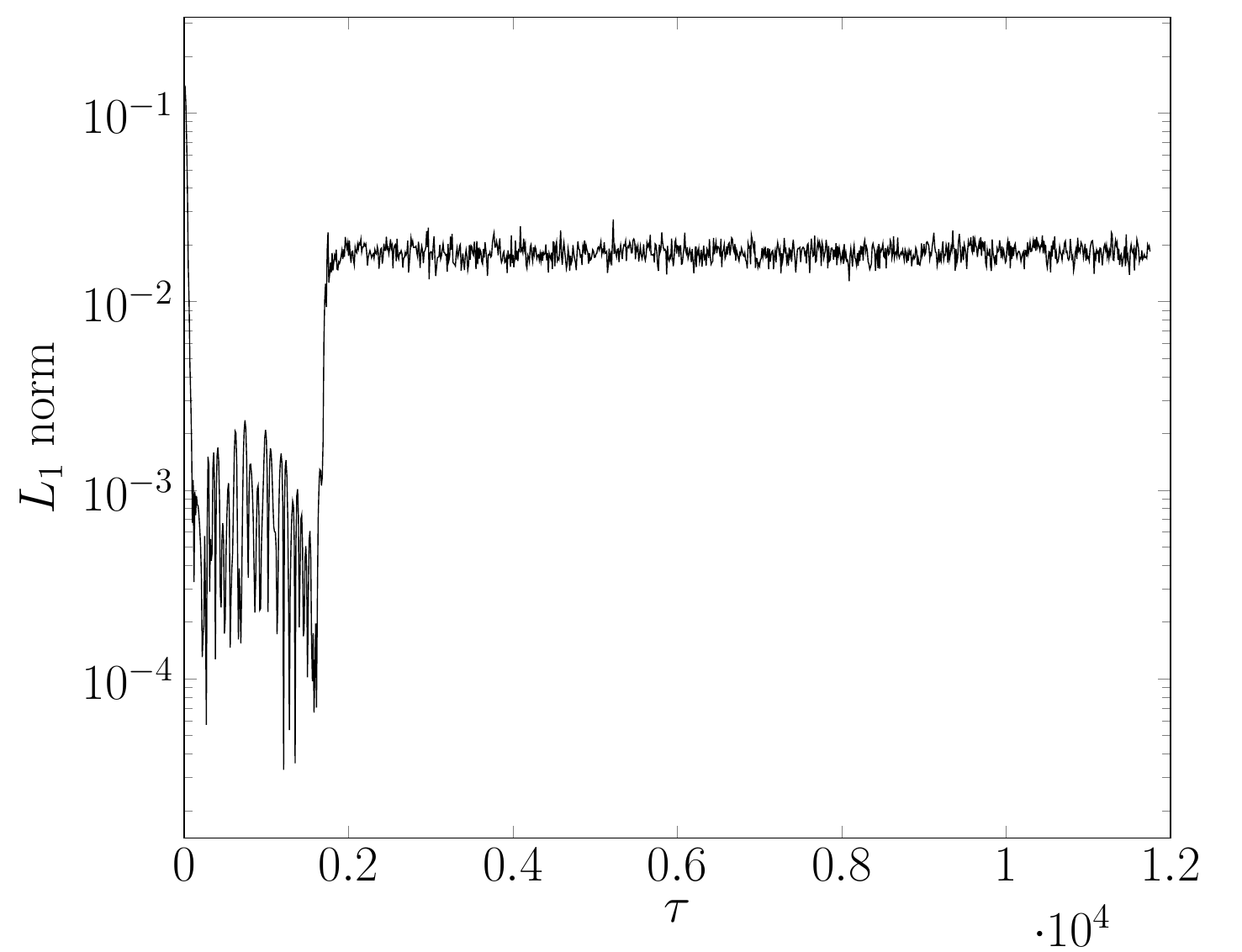}
        		\caption{$L_1$ norm evolution}
        		\label{fig:chaos2b}
    	\end{subfigure}
    	\vspace{2mm}
    	\caption{Numerical solution for a 2D anisotropic DKS equation - Chaotic semi-organized 
    	oscillatory behavior, with $\alpha = 0.05$ and $\theta = 30^\circ$.}
    	\label{fig:chaos2}
	\end{figure*}

	Figure \ref{fig:chaos2} reveals the numerical solution for $\alpha = 0.05$. The initial
	condition was the same as the previous case ($q_c = 1.7181 \cdot 10^{-1}$) . A 
	spatiotemporal chaotic cellular structure is obtained for late time,  which can be seen in 
	Fig. \ref{fig:chaos1a} for $\tau = 11,750$. In comparison with the undamped structure,
	the late time pattern for $\alpha = 0.05$ is much more organized, with a smaller variation
	of cell sizes and shape, where some of them approach the critical $\lambda_c$ width. The 
	$L_1$ norm evolution (Fig. \ref{fig:chaos2b}) shows that a strongly oscillatory state is
	reached about $\tau = 2,000$, where $L_1$ starts fluctuating around $L_1 = 0.02$. These
	intense dynamics differ from the undamped case: even though the structure is more 
	organized, it keeps changing at a constant rate for an undefined period of time.

	Both cases differ from the simulation with $\alpha = 0.15$, where the damping effect is 
	sufficient for an ordered and quick reorganization of the structure after a $L_1$ peak. 
	In the $\alpha = 0.05$ case, the damping is not high enough to allow the microstructure to 
	reorganize itself into a perfectly ordered hexagonal state, and it keeps chaotically 
	oscillating  around the peak $L_1$ value. Another observation is made towards the 
	obtained height values: analyzing $\alpha = 0$ case, the mean height of the 
	surface falls continuously with time, while maintaining the distance $\bar{h}_{dif}$ 
	between the minimum and maximum points around $\bar{h}_{dif} = 6.4$. However, 
	for $\alpha = 0.05$, the mean height remains approximately constant, oscillating from -0.5 to 
	-3.5 ($\bar{h}_{dif} \approx 3.0$), for an undefined time. In comparison, for the steady state 
	obtained with a damping $\alpha = 0.15$ and $\vec{q} = q_{0}\vec{1}_{X}$,  the maximum 
	and minimum height values were, respectively,  -0.03 and -0.81 ($\bar{h}_{dif} = 0.78$).

	The practical significance of those results is that the acquired structures might depend on
	the duration of the irradiation. For a short time duration sputtering in a system with
	$\alpha = 0.15$, we may have only ripple formation, while hexagons shall emerge for a
	longer duration of the experiment. In contrast, for a system with $\alpha = 0$ or $0.05$, 
	the final structure will depend on time, since it keeps changing with the irradiation 
	duration. We note that due to an unstable initial growth, the case without damping
    does not present proper physical results, since the magnitude of $h$ grew
    surpassing reasonable boundaries. The curve $L_1$ for $\alpha = 0$ seems to decay
    in the long time, and further investigations could be made. Still, the present
    results reinforce the importance of the linear damping in the modelling of
    sputtering.


	\subsection{Anisotropy and angle of incidence}

\begin{figure*}[htbp!]
		\centering
		\includegraphics[width=0.55\linewidth]{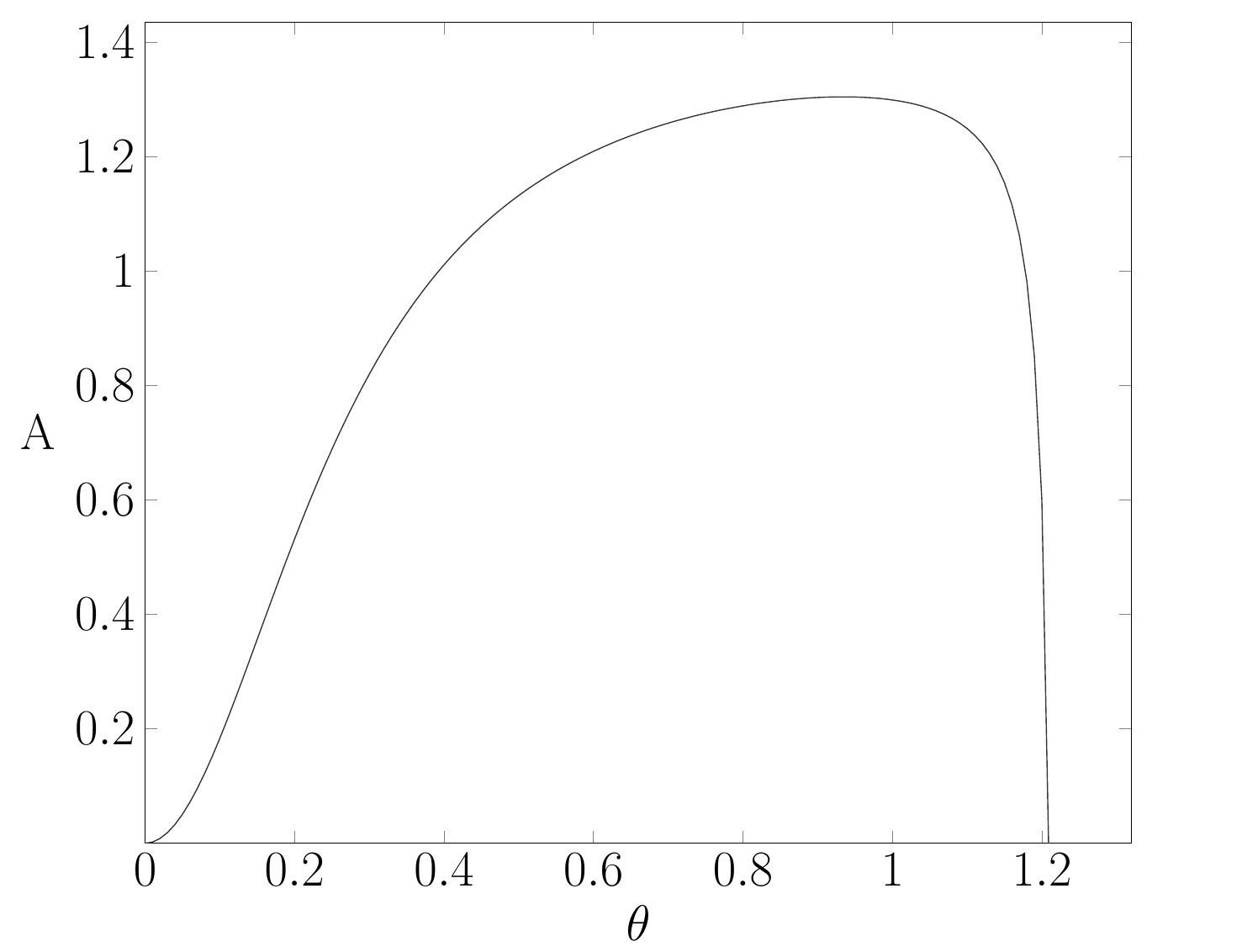}
		\caption{Relative anisotropy $A$ versus $\theta$ for $\bar{K} = 5$ and $a_\eta = 4$ .}
		\label{fig:anisotropy}
	\end{figure*}

	From our previous weakly nonlinear analysis, we are able to investigate 
	the parameters involved in the relative anisotropy of the resulting patterns. 
	The relative anisotropy is studied by $A = 
	\Omega/\bar{K} q_c^4$, where $q_c$ is the critical wavenumber obtained from our 
	previous linear stability analysis, and $\Omega = \frac{3}{4}(|\mu| - |\nu|) q_c^2$. 
	Figure \ref{fig:anisotropy} plots the relative anisotropy $A$ versus $\theta$ for 
	$\bar{K} = 5$ and $a_\eta = 4$. From the plot we observe that $A$ for $\theta = 0.5236$ 
	($30^\circ$) is approximately 1.15, which was the value used up to this moment.
	The following results investigate the effect of varying the angle of incidence in the 
	resulting pattern, motivated by this behavior of the relative anisotropy.
	
	One region that we are particularly interested in is near $\theta = 1.2$, since $\bar\nu_x$ and 
	$\bar\nu_y$ may present opposite signs with similar absolute value. This observation implies that 
	there will be one direction for the unstable modes where the nonlinear terms ($\bar\nu_x$ and 
	$\bar\nu_y$ from Eq. \ref{eq:KS2}) will compensate each other, as studied in the work of Rost and 
	Krug \cite{rost1995anisotropic} (without damping). For such simulation, we set  $\theta$ to 1.1549 
	($66.17^\circ$), which means that we will have the following anisotropy coefficients: 
	$\bar\nu_x = 0.0658$ and $\bar\nu_y = -0.0659$. The damping coefficient is set to $\alpha = 0.1$.

\begin{figure*}[h!]
	\centering
	   \begin{subfigure}[t]{0.48\textwidth}
   		\hspace{1mm}
       		\includegraphics[width=1.0\linewidth]
        		{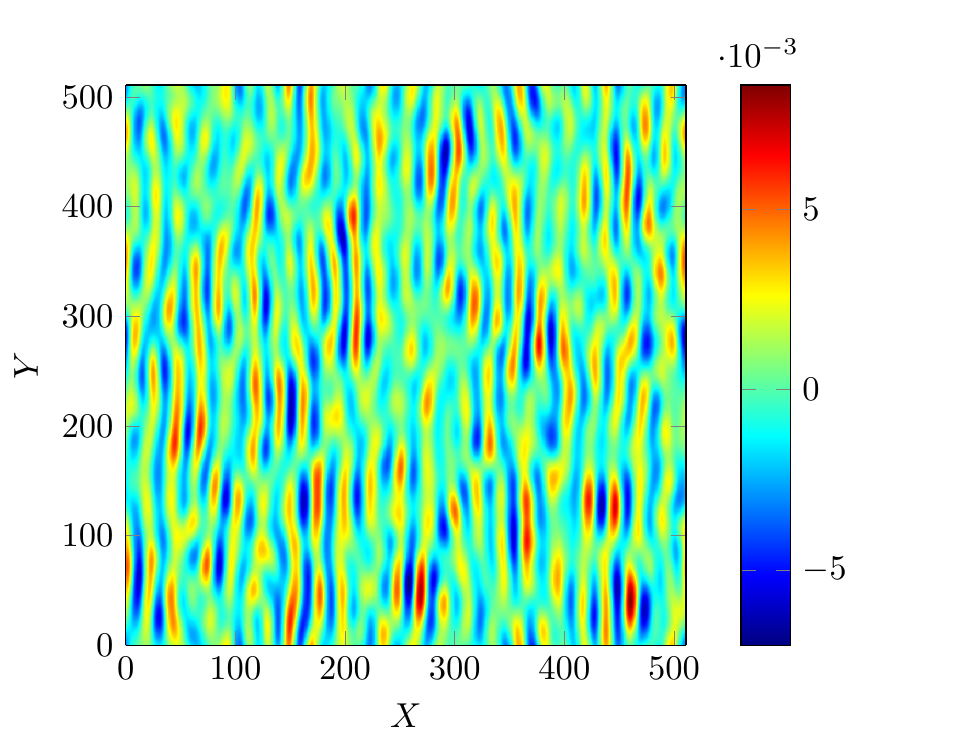}
       		\caption{$\tau = 18$}
    	\end{subfigure}
    	\hspace{1mm}
    	\begin{subfigure}[t]{0.48\textwidth}
    		\hspace{1mm}
        		\includegraphics[width=1.0\linewidth]
        		{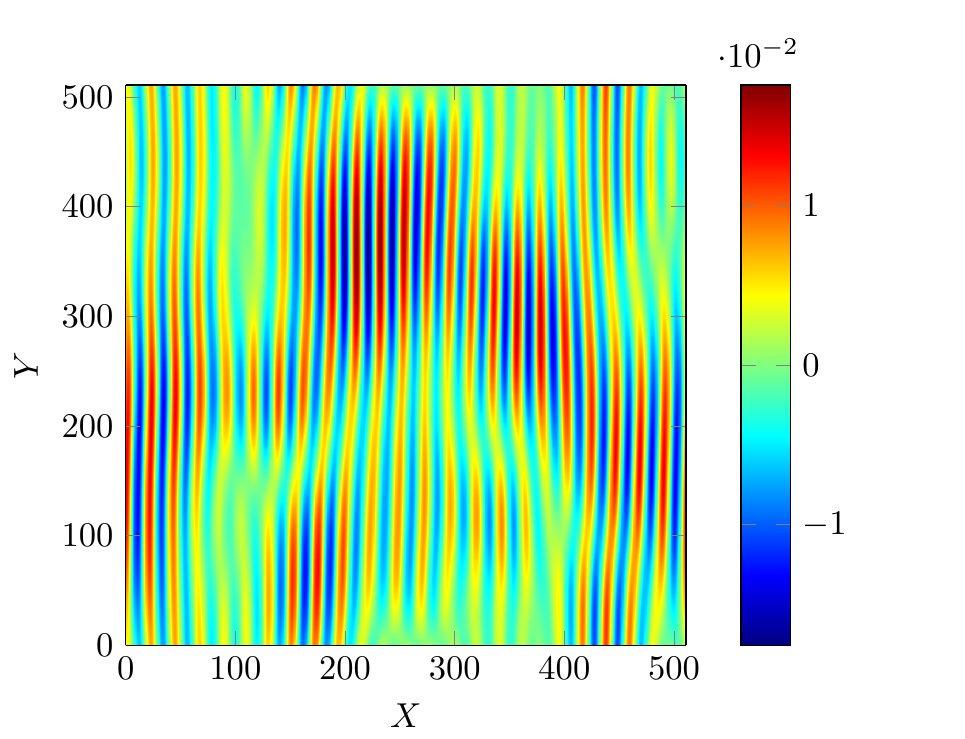}
        		\caption{$\tau = 180$}
	\end{subfigure}
	\caption{Simulation results for a system 512 $\times$ 512 with an angle $\theta = 
        66.17^\circ$. The nonlinearities ``cancel'' each other and a well defined
        direction arises from the unstable mode. This is made clear by the selection
        of the $\vec{1}_x$  direction for $\tau = 180$.}
	\label{fig:cancel}
	\end{figure*}

	Figure \ref{fig:cancel} shows that the nonlinearities compensate each other 
	when the system remains aligned with  the $\vec{1}_x$ direction, even for unstable 
	modes. The nanostructure obtained for $\tau = 180$ is still irregular in terms of the ripple 
	behavior, but the direction of preference is clear. There are approximately 23 wavelengths 
	in the domain, which is less than the critical number of wavelengths from the linear 
	stability analysis. In this case, we did not obtain a stationary structure, since the irregular 
	ripple morphology keeps evolving, although the pattern's preferred direction remains the 
	same.

\section{Conclusions}

	In the present work we have developed a finite-difference time splitting scheme to
	solve an anisotropic  damped Kuramoto-Sivashinsky equation, which comes from the
	continuum theory and is an option to describe a surface eroded by ion bombardment.
	We dealt with realistic coefficients, based on physical values typically found
    in the literature. The dimensionless analysis was performed for a high
    temperature scenario ($T = 500K$).
	
	The MMS was employed for code verification, and a second-order convergence was 
	detected for coarser meshes comparison, while results between first and
    second-order convergence came up for more refined meshes, suggesting a possible
    issue with the manufactured solution stability. Regarding the scheme's stability,
    the tests revealed that for $\Delta \tau \leq 2.0$, the numerical scheme was
    sufficiently stable with a grid spacing $\Delta X = 1.0$. 
	
	Spatiotemporal chaotic structures appeared for the undamped case, whose dynamics
    fell continuously for the long time. A chaotic oscillatory pattern rose from the
    simulation with $\alpha = 0.05$, reaching a better ordered structure than the one
    for the undamped result, while maintaining a pattern under oscillatory evolution
    after the emergence of the hexagonal modes. Defectless hexagonal periodic
    structures  were obtained  for higher values of the damping coefficient, with an
    angle of incidence $\theta = 30^\circ$. Although its physical origin has been
    questioned in the literature \cite{bradley2010spontaneous}, the damping term is an
	essential ingredient of the present model to obtain the desired nanohole pattern.

	Regarding the effect of preexisting patterns over the steady state in a system
    with higher damping coefficient ($\alpha \geq 0.1$), no appreciable difference 
	was observed in the final morphology born from random or monomodal initial 
	patterns. For all considered initial structures, we have arrived in the
    aforementioned defectless nanohole pattern.  However, the evolution of
    $\vec{1}_X$ monomodal patterns was significantly different when compared to the
    others, since the former was the only scenario where the nanostructuration of a
    1D pattern  with a wavenumber near $q_c$ was observed before the emergence of
    hexagonal modes.

	Based on the previous work of Rost and Krug, we investigated a case where the 
	nonlinearities compensate each other. In their model, there was a subtle balance
    between the anisotropy of both the linear terms and nonlinear terms, leading to
    nonintuitive results, suggesting that a numerical analysis is essential. In this
    paper, the nonlinear compensation was achieved using $\theta = 66.17^\circ$, and
    an irregular oscillatory ripple structure with a clear orientation in the
    $\vec{1}_X$ direction was obtained.
	
	In summary, the studied model equation leads to ripple morphologies and nanohole
    patterns, such that the prevailing structure is sensitive to the anisotropy of
    the system, while unaffected by changes in the initial pattern. The hexagonal
    structures are equally attained through experiments, and the present results
    reinforce the role of the irradiation duration for the acquired surface
    morphology.


 \section*{Acknowledgement}

    Eduardo Vitral acknowledges a fellowship from the Coordination for the
    Improvement of Higher Education-CAPES (Brazil). A FAPERJ Senior Researcher
    Fellowship is acknowledged by Jos\'e Pontes. Gustavo Anjos is granted by
    Science Without Borders/Young Talent Attraction program (CAPES).



\bibliography{sputDKS}

\begin{thebibliography}{10}
\expandafter\ifx\csname url\endcsname\relax
  \def\url#1{\texttt{#1}}\fi
\expandafter\ifx\csname urlprefix\endcsname\relax\def\urlprefix{URL }\fi
\expandafter\ifx\csname href\endcsname\relax
  \def\href#1#2{#2} \def\path#1{#1}\fi

\bibitem{chason2010spontaneous}
E.~Chason, W.~L. Chan, Spontaneous patterning of surfaces by low-energy ion
  beams, in: Materials Science with Ion Beams, Springer, 2010, pp. 53--71.

\bibitem{gago2006temperature}
R.~Gago, L.~V{\'a}zquez, O.~Plantevin, J.~S{\'a}nchez-Garc{\'\i}a, M.~Varela,
  M.~Ballesteros, J.~Albella, T.~Metzger, Temperature influence on the
  production of nanodot patterns by ion beam sputtering of {S}i (001), Physical
  Review B 73~(15) (2006) 155414.

\bibitem{wei2009self}
Q.~Wei, X.~Zhou, B.~Joshi, Y.~Chen, K.-D. Li, Q.~Wei, K.~Sun, L.~Wang,
  Self-assembly of ordered semiconductor nanoholes by ion beam sputtering,
  Advanced Materials 21~(28) (2009) 2865--2869.

\bibitem{valbusa2002nanostructuring}
U.~Valbusa, C.~Boragno, F.~B. de~Mongeot, Nanostructuring surfaces by ion
  sputtering, Journal of Physics: Condensed Matter 14~(35) (2002) 8153.

\bibitem{mollick2014formation}
S.~A. Mollick, D.~Ghose, B.~Satpati, Formation of {A}u--{G}e nanodots by {A}u-
  ion sputtering of {G}e, Vacuum 99 (2014) 289--293.

\bibitem{makeev2002morphology}
M.~A. Makeev, R.~Cuerno, A.-L. Barab{\'a}si, Morphology of ion-sputtered
  surfaces, Nuclear Instruments and Methods in Physics Research Section B: Beam
  Interactions with Materials and Atoms 197~(3) (2002) 185--227.

\bibitem{rost1995anisotropic}
M.~Rost, J.~Krug, Anisotropic {K}uramoto-{S}ivashinsky equation for surface
  growth and erosion, Physical {R}eview {L}etters 75~(21) (1995) 3894.

\bibitem{paniconi1997stationary}
M.~Paniconi, K.~Elder, Stationary, dynamical, and chaotic states of the
  two-dimensional damped {K}uramoto-{S}ivashinsky equation, Physical Review E
  56~(3) (1997) 2713.

\bibitem{facsko2004dissipative}
S.~Facsko, T.~Bobek, A.~Stahl, H.~Kurz, T.~Dekorsy, Dissipative continuum model
  for self-organized pattern formation during ion-beam erosion, Physical Review
  B 69~(15) (2004) 153412.

\bibitem{keller2010ion}
A.~Keller, S.~Facsko, Ion-induced nanoscale ripple patterns on {S}i surfaces:
  theory and experiment, Materials 3~(10) (2010) 4811--4841.

\bibitem{christov1997implicit}
C.~Christov, J.~Pontes, D.~Walgraef, M.~G. Velarde, Implicit time splitting for
  fourth-order parabolic equations, Computer {M}ethods in {A}pplied {M}echanics
  and {E}ngineering 148~(3) (1997) 209--224.

\bibitem{christov2002numerical}
C.~Christov, J.~Pontes, Numerical scheme for {S}wift-{H}ohenberg equation with
  strict implementation of {L}yapunov functional, Mathematical and {C}omputer
  {M}odelling 35~(1) (2002) 87--99.

\bibitem{sigmund1969theory}
P.~Sigmund, Theory of {S}puttering. i. {S}puttering yield of amorphous and
  polycrystalline targets, Physical Review 184 (1969) 383--416.

\bibitem{bradley1988theory}
R.~M. Bradley, J.~M. Harper, Theory of ripple topography induced by ion
  bombardment, Journal of Vacuum Science \& Technology A 6~(4) (1988)
  2390--2395.

\bibitem{cuerno1995dynamic}
R.~Cuerno, A.-L. Barab{\'a}si, Dynamic scaling of ion-sputtered surfaces,
  Physical {R}eview {L}etters 74~(23) (1995) 4746.

\bibitem{bradley2011redeposition}
R.~M. Bradley, Redeposition of sputtered material is a nonlinear effect,
  Physical Review B 83~(7) (2011) 075404.

\bibitem{douglas1956numerical}
J.~Douglas, H.~H. Rachford, On the numerical solution of heat conduction
  problems in two and three space variables, Transactions of the American
  mathematical Society (1956) 421--439.

\bibitem{yanenko1971method}
N.~N. Yanenko, The method of fractional steps, Springer, 1971.

\bibitem{malaya2013masa}
N.~Malaya, K.~C. Estacio-Hiroms, R.~H. Stogner, K.~W. Schulz, P.~T. Bauman,
  G.~F. Carey, {MASA}: a library for verification using manufactured and
  analytical solutions, Engineering with Computers 29~(4) (2013) 487--496.

\bibitem{roy2005review}
C.~J. Roy, Review of code and solution verification procedures for
  computational simulation, Journal of Computational Physics 205~(1) (2005)
  131--156.

\bibitem{ghoniem2008instabilities}
N.~Ghoniem, D.~Walgraef, Instabilities and Self-organization in Materials,
  Oxford Univ. Press, 2008.

\bibitem{walgraefProposal}
D.~Walgraef, Nano-patterning of surfaces by ion sputtering: A proposal for a
  numerical study of the effect of preexisting patterns, working notes,
  unpublished.

\bibitem{manneville1990dissipative}
P.~Manneville, Dissipative structures and weak turbulence, San Diego, CA (USA);
  Academic Press Inc., 1990.

\bibitem{roache2002code}
P.~J. Roache, Code verification by the method of manufactured solutions,
  Journal of Fluids Engineering 124~(1) (2002) 4--10.

\bibitem{bradley2010spontaneous}
R.~M. Bradley, P.~D. Shipman, Spontaneous pattern formation induced by ion
  bombardment of binary compounds, Physical {R}eview {L}etters 105~(14) (2010)
  145501.

\end{thebibliography}

\end{document}